%% file: aMCfast2.tex
\documentclass[11pt,a4paper]{article}
\usepackage{graphicx}
\usepackage{afterpage}
\usepackage{epsfig,cite}
\usepackage{booktabs,multirow,tabularx}
\usepackage{amssymb}
\usepackage{amsmath}
\usepackage{dsfont}
\usepackage{multirow}
\usepackage{url,hyperref}
\usepackage[Q=yes,pverb-linebreak=no]{examplep}

\usepackage{color}
\definecolor{comment}{rgb}{0,0.3,0}
\definecolor{identifier}{rgb}{0.0,0,0.3}

\usepackage{listings}

\usepackage{url}
\usepackage{hyperref}

\def\beq{\begin{equation}}
\def\beqn{\begin{eqnarray}}
\def\eeq{\end{equation}}
\def\eeqn{\end{eqnarray}}
\def\abs#1{\left|#1\right|}
\def\floor#1{\left\lfloor #1\right\rfloor}

\newcommand\mydot{\!\cdot\!}

\newcommand\muFak{\mu_{{\sss F};k}^{(\alpha)}}
\newcommand\muRak{\mu_{{\sss R};k}^{(\alpha)}}
\newcommand\tauak{\tau_k^{(\alpha)}}
\newcommand\aNLO{{\sc\small MadGraph5\_aMC@NLO}}
\newcommand\amcfast{{\sc\small aMCfast}}
\newcommand\applgr{{\sc\small APPLgrid}}
\newcommand\fastnlo{{\sc\small FastNLO}}

\newcommand\db{\bar{d}}
\newcommand\ub{\bar{u}}
\newcommand\bs{\bar{s}}
\newcommand\cb{\bar{c}}
\newcommand\bb{\bar{b}}

\newcommand\stepf{\Theta}

\newcommand\conf{{\cal K}}

\newcommand\confnpoak{\conf_{n+1;k}^{(\alpha)}}

\newcommand\confnpoSk{\conf_{n+1;k}^{(S)}}

\newcommand\wka{\Xi_k^{(\alpha)}}

\newcommand\stepkha{\stepf_k^{(h,\alpha)}}

\newcommand\as{\alpha_{\sss S}}
\newcommand\aem{\alpha}

\newcommand\gs{g_{\sss S}}
\newcommand\hgs{\hat{g}_{\sss S}}

\newcommand\NF{N_{\sss F}}

\newcommand\MadFKS{{\sc\small MadFKS}}
\newcommand\CutTools{{\sc\small CutTools}}
\newcommand\OL{{\sc\small OpenLoops}}
\newcommand\MadLoop{{\sc\small MadLoop}}

\newcommand\xiF{\xi_{\sss F}}
\newcommand\xiR{\xi_{\sss R}}
\newcommand\muF{\mu_{\sss F}}
\newcommand\muR{\mu_{\sss R}}
\newcommand\muFa{\mu_{\sss F}^{(\alpha)}}
\newcommand\muRa{\mu_{\sss R}^{(\alpha)}}

\newcommand\muFaoQt{\left(\!\frac{\muFa}{\QESa}\!\right)^2}
\newcommand\muRaoQt{\left(\!\frac{\muRa}{\QESa}\!\right)^2}
\newcommand\muFakoQt{\left(\!\frac{\muFak}{\QESa}\!\right)^2}
\newcommand\muRakoQt{\left(\!\frac{\muRak}{\QESa}\!\right)^2}
\newcommand\muFoQt{\left(\!\frac{\muF}{\QESa}\!\right)^2}
\newcommand\muRoQt{\left(\!\frac{\muR}{\QESa}\!\right)^2}

\newcommand\QESa{Q}

\newcommand\fo{f_1}
\newcommand\ft{f_2}

\newcommand\fr{f_r}
\newcommand\fs{f_s}

\newcommand\luml{{\cal F}^{(l)}}
\newcommand\hluml{\widehat{\cal F}^{(l)}}
\newcommand\xo{x_1}
\newcommand\xt{x_2}
\newcommand\yo{y_1}
\newcommand\yt{y_2}
\newcommand\xoa{x_1^{(\alpha)}}
\newcommand\xta{x_2^{(\alpha)}}

\newcommand\xoak{x_{1;k}^{(\alpha)}}
\newcommand\xtak{x_{2;k}^{(\alpha)}}
\newcommand\xiak{x_{i;k}^{(\alpha)}}
\newcommand\yoak{y_{1;k}^{(\alpha)}}
\newcommand\ytak{y_{2;k}^{(\alpha)}}
\newcommand\yiak{y_{i;k}^{(\alpha)}}

\newcommand\Wa{W^{(\alpha)}}

\newcommand\Walq{W_{lq}^{(\alpha)}}

\newcommand\Warsq{W_{rsq}^{(\alpha)}}
\newcommand\Warsqp{W_{r^\prime s^\prime q^\prime}^{(\alpha)}}
\newcommand\hWaz{\widehat{W}^{(\alpha)}_0}
\newcommand\hWaF{\widehat{W}^{(\alpha)}_{\sss F}}
\newcommand\hWaR{\widehat{W}^{(\alpha)}_{\sss R}}
\newcommand\hWabe{\widehat{W}^{(\alpha)}_\beta}
\newcommand\hWazk{\widehat{W}^{(\alpha)}_{0;k}}
\newcommand\hWaFk{\widehat{W}^{(\alpha)}_{{\sss F};k}}
\newcommand\hWaRk{\widehat{W}^{(\alpha)}_{{\sss R};k}}
\newcommand\hWabek{\widehat{W}^{(\alpha)}_{\beta;k}}

\newcommand\hWabelqk{\widehat{W}^{(\alpha)}_{\beta,lq;k}}

\newcommand\hWz{\widehat{W}_0}
\newcommand\hWF{\widehat{W}_{\sss F}}
\newcommand\hWR{\widehat{W}_{\sss R}}
\newcommand\hWB{\widehat{W}_{\sss B}}

\newcommand\hWBak{\widehat{W}^{(\alpha)}_{{\sss B};k}}

\newcommand\sigmaNLO{\sigma^{\sss\rm (NLO)}}
\newcommand\sigmaNLOa{\sigma^{{\sss (\rm NLO},\alpha{\sss )}}}

\newcommand{\pt}{p_{\sss T}}
\newcommand{\Et}{E_{\sss T}}
\newcommand{\kt}{k_{\sss T}}
\newcommand{\meas}{\chi}

\newcommand{\bt}{\bar{t}}
\newcommand\prompt{{\tt MG5\_aMC>}}

\newcommand{\be}{\begin{equation}}
\newcommand{\ee}{\end{equation}}
\newcommand{\bea}{\begin{eqnarray}}
\newcommand{\eea}{\end{eqnarray}}
\newcommand{\bi}{\begin{itemize}}
\newcommand{\ei}{\end{itemize}}
\newcommand{\ben}{\begin{enumerate}}
\newcommand{\een}{\end{enumerate}}

\newcommand{\lp}{\left(}
\newcommand{\rp}{\right)}

\def\frac#1#2{{{#1}\over {#2}}}
\def\gsim{\mathrel{\rlap{\lower4pt\hbox{\hskip1pt$\sim$}}
    \raise1pt\hbox{$>$}}}         
\def\lsim{\mathrel{\rlap{\lower4pt\hbox{\hskip1pt$\sim$}}
    \raise1pt\hbox{$<$}}}         

\newcommand{\draft}[1]{}

\def\beq{\begin{equation}}  
\def\eeq{\end{equation}}

\newcommand\sss{\scriptscriptstyle}

\def\beq{\begin{equation}}
\def\eeq{\end{equation}}
\def\beqn{\begin{eqnarray}}
\def\eeqn{\end{eqnarray}}
\def\abs#1{\left|#1\right|}

\begin{document}

\vspace{-2.0cm}
\begin{flushright}
CERN-PH-TH/2014-009\\
OUTP-14-10p \\
\end{flushright}

\begin{center}
{\Large \bf {\sc aMCfast}: automation of fast NLO computations for PDF fits}
\vspace{.7cm}

Valerio~Bertone$^{1}$,
Rikkert Frederix$^{1}$,
Stefano~Frixione$^{1}$,
Juan~Rojo$^{1,2}$, 
and Mark~Sutton$^{3}$

\vspace{.3cm}
{\it ~$^1$ PH Department, TH Unit, CERN, CH-1211 Geneva 23, Switzerland \\
~$^2$ Rudolf Peierls Centre for Theoretical Physics, 1 Keble Road,\\ 
University of Oxford, OX1 3NP Oxford, United Kingdom\\
~$^3$ Department of Physics and Astronomy, University of Sussex, 
United Kingdom\\}
\end{center}   

\vspace{0.1cm}

\begin{center}
{\bf \large Abstract}
\end{center}
We present the interface between \aNLO, a self-contained program that 
calculates cross sections up to next-to-leading order accuracy in an automated
manner, and \applgr, a code that parametrises such cross sections in the 
form of look-up tables which can be used for the fast computations needed 
in the context of PDF fits. The main characteristic of this interface,
which we dub \amcfast, is its being fully automated as well, which 
removes the need to extract manually the process-specific information 
for additional physics processes, as is the case with other matrix-element 
calculators, and renders it straightforward to include any new process 
in the PDF fits. We demonstrate this by 
studying several cases which are easily measured at the LHC, have a good
constraining power on PDFs, and some of which were previously unavailable 
in the form of a fast interface. 

\clearpage

\tableofcontents

\clearpage

\section{Introduction}\label{sec:intro}

The accurate determination of the Parton Distribution Functions (PDFs) of the
proton~\cite{Forte:2013wc,Ball:2012wy,Perez:2012um,Forte:2010dt,DeRoeck:2011na}
is one of the most important tasks for precision phenomenology at the
Large Hadron Collider (LHC).  PDFs are a dominant source of theoretical
uncertainty in the predictions for Higgs boson production, where the errors
that affect them degrade the
accuracy of the Higgs characterization in terms of couplings and branching
fractions~\cite{Dittmaier:2011ti}; they induce large uncertainties in
the cross sections for processes with very massive new-physics
particles~\cite{AbelleiraFernandez:2012ty};
and they substantially affect Standard Model (SM) precision measurements such 
as those of the mass of the $W$ boson~\cite{Bozzi:2011ww} and of the effective
lepton mixing angle $\sin^2\theta^l_{\rm eff}$~\cite{Rojo:2013wba}.  
For these reasons, an active program towards better PDFs is being carried 
out by different groups~\cite{Martin:2009iq,Ball:2010de,Gao:2013xoa,
Alekhin:2013nda,CooperSarkar:2011aa}, which emphasise the use of new 
experimental inputs, more accurate theoretical calculations,
and improved fitting methodology.

Modern global PDF analyses are based on a variety of hard-scattering data 
such as deep-inelastic scattering (DIS) structure functions at fixed-target
experiments, lepton-proton cross sections from the HERA collider, and
inclusive $W$, $Z$, and jet production at hadron colliders.  Since the
beginning of the LHC data taking, these and many other processes for which
measurements have become available and can be used to constrain PDFs.  
For example, LHC data that might provide information on PDFs 
encompass inclusive electroweak vector boson production~\cite{Aad:2011dm,
Chatrchyan:2013mza}, inclusive jet and dijet production~\cite{Aad:2011fc,
Chatrchyan:2012bja,Aad:2013tea,Aad:2013lpa,Watt:2013oha},
direct photon production~\cite{d'Enterria:2012yj}, top quark pair production
cross sections~\cite{Czakon:2013tha}, $W$ production in association with charm
quarks~\cite{Chatrchyan:2013uja,Chatrchyan:2013mza}, low and high-mass
Drell-Yan production~\cite{Chatrchyan:2013tia}, the $W$ and $Z$ bosons 
large-$\pt$ distributions and their ratios~\cite{Malik:2013kba}, high-mass $W$
production, and single-top production, as well as ratios of cross sections
measured at different center-of-mass energies~\cite{Mangano:2012mh,
Aad:2013lpa}.

It is thus clear that a wide variety of high-quality measurements sensitive to
PDFs are already available; more data will follow in the coming years.
Therefore, in order to improve the accuracy of PDF determinations, it is
essential to include in PDF fits as many of these measurements as possible, in
order to constrain different combinations of PDFs in a wide range of
Bjorken $x$'s.  The most serious difficulty in doing so is due to the fact that
next-to-leading order (NLO) QCD calculations of hadron collider processes
with realistic acceptance cuts are much slower than what is
needed in the iterative PDF fitting procedure, which requires computing the
theoretical predictions a very large number of times.  In order to bypass this
problem, a popular solution in the past has been that of performing leading
order (LO) computations, supplemented by bin-by-bin $K$ factors.
Unfortunately, such a solution is not sufficiently accurate to match the
precision of present and future LHC data; in particular, it has the very
undesirable feature of neglecting the combinations of initial-state partons
that do not appear at the LO.

In order to overcome this problem, several solutions have been proposed. The
underlying idea common to all these approaches is:  interpolating the
PDFs in the $\lp x,Q^2\rp$-plane with some suitable polynomial basis; 
precomputing the hadronic cross section by using the basis members as input;
and finally reconstructing the original calculation with the numerical
convolution of the precomputed cross sections and the actual PDFs. Note that
the information about the latter is only required at a {\em finite} number of 
points $(x_i,Q^2_j)$, which are called the interpolating-grid nodes.  
Therefore, the time-consuming task of the precomputation of the cross section
with basis members is performed only once, and the reconstruction of the full
result associated with arbitrary PDFs is extremely fast.

The strategy sketched above, which is closely related to the one adopted by
$x$-space PDF evolution codes~\cite{Salam:2008qg,Botje:2010ay,Bertone:2013vaa}, 
is what underpins the two best-known fast interpolators of NLO QCD
cross sections, \fastnlo~\cite{Kluge:2006xs,Wobisch:2011ij} and
\applgr~\cite{Carli:2010rw}. \fastnlo\ is interfaced to the jet-production code
{\sc\small NLOjet++}~\cite{Nagy:2001fj}, and is thus able to provide fast
computations of multijet production at lepton-proton and hadron-hadron 
colliders\footnote{Recently, the \fastnlo\ interface has been generalised 
to processes other than jet production, for instance to the approximate 
NNLO calculation of differential distributions in top-pair 
production of ref.~\cite{Guzzi:2014wia}.}.
\applgr\ is interfaced to various programs, including
{\sc\small NLOjet++} and {\sc\small MCFM}~\cite{Campbell:2002tg}.
The main drawback of these tools, namely that of being capable of
handling a relatively small number of processes, is a consequence
of the fact that adding new ones requires an ad-hoc procedure,
which is rather time-consuming and error-prone. This is also the
principal reason why they are only interfaced to calculations which are
of NLO (and, in the one case mentioned above, NNLO)
in QCD, but neither to (N)LO results matched to parton showers,
nor to NLO results in the electroweak theory.

The goal of this work is that of solving all of these problems
in a general manner, which is possible thanks to the fact that
NLO(+PS) calculations can now be routinely done by means of
automated codes. In fact, among the many features of such codes
there are two which are directly relevant to the problems at hand:
firstly, given a sufficient amount of CPU power the cross section
for any process, however complicated, can be computed; and secondly,
the way in which these cross sections are handled and the form in
which they are written are completely standardised. This is what
renders it possible to construct a generic and automated interface
between an automated cross section calculator and a fast interpolator.
The main result of this paper is the construction of such an automated
interface, that we call \amcfast, and which bridges the automated
cross section calculator \aNLO~\cite{Alwall:2014hca} with the
fast interpolator \applgr.
Thus, the chain \aNLO\ -- \amcfast\ -- \applgr\ will allow one
to include, in a straighforward manner, any present or future LHC 
measurement in an NLO global PDF analysis. We point out that
a strategy analogous to that pursued in the present paper has motivated
the recent construction of {\sc\small MCgrid}~\cite{DelDebbio:2013kxa},
whereby \applgr\ has been interfaced to 
{\sc\small Rivet}~\cite{Buckley:2010ar}.

We remind the reader that \aNLO\ contains all ingredients relevant
to the computations of LO and NLO cross sections, with or without
matching to parton showers. NLO results not matched to parton showers
(called fNLO~\cite{Alwall:2014hca}) are obtained by adopting the
FKS method~\cite{Frixione:1995ms,Frixione:1997np} for the subtraction 
of the singularities of the real-emission matrix elements (automated 
in the module \MadFKS~\cite{Frederix:2009yq}), and the OPP
integral-reduction procedure~\cite{Ossola:2006us} for the computation 
of the one-loop matrix elements (automated in the module 
\MadLoop~\cite{Hirschi:2011pa}, which makes use of
\CutTools~\cite{Ossola:2007ax} and of an in-house implementation 
of the optimisations proposed in ref.~\cite{Cascioli:2011va} (\OL)).
Matching with parton showers is achieved
by means of the MC@NLO formalism~\cite{Frixione:2002ik}.
In the present public version, \aNLO\ is restricted to computing
NLO QCD corrections to SM processes; however, as discussed in
ref.~\cite{Alwall:2014hca}, all obstacles that enforce such a limitation
have been cleared, paving the way to higher-order calculations
in the context of arbitrary renormalisable theories in the near future.
As far as \applgr\ is concerned, only a few high-level routines
have been extended in view of its interface with \amcfast. All of
these modifications are of a technical character, except one which
is related to the computation of the factorisation and renormalisation
scale dependences, as we shall explain in more detail later.

The scope of this paper is that of fNLO QCD computations.
Alternatively, in \amcfast\ the specific nature of the perturbative 
expansion is used only in a rather trivial way, since it determines
a number of interpolating grids and their linear combination that
defines the physical cross sections. Therefore, when e.g.~electroweak
corrections will become publicly available in \aNLO, that feature will 
be immediately inherited by \amcfast\ through some straighforward 
generalisation. Furthermore, what is done here will allow one, with only 
a few minimal extensions, to construct fast interfaces to NLO+PS 
predictions. This is expected to have several beneficial effects in the 
context of PDF fits: a closer connection to the experimentally 
accessible observables, a wider range of data that can be used to 
constrain PDFs, and the possibility of eventually extract 
PDFs that are specifically tailored to their use with 
NLO event generators.

This paper is organised as follows:  in sect.~\ref{sec:amcfast} 
we give a short introduction to the interpolating-grid techniques employed 
here, review the computation of cross sections in \aNLO, and 
discuss how the separation between PDFs, partonic cross
sections, and $\as$ dependence can be exploited to construct a fast
interface using the \applgr\ routines.  The flexibility of \amcfast\
is illustrated in sect.~\ref{sec:pheno}, where we present 
results for many relevant LHC processes, some of which are obtained for 
the first time with a fast NLO interface, and whose numerical performance
and accuracy are analysed. Finally, in sect.~\ref{sec:outlook} we 
summarise our findings and briefly discuss the plans
for the future developments of \amcfast. An appendix reports 
some additional information.

\section{Automation of fast NLO computations\label{sec:amcfast}}
In this section we first outline the basics of an interpolation
technique based on the expansion of a given function onto a basis 
of polynomials; we then discuss how short-distance cross
sections can be represented, in the most general manner, in
terms of interpolating grids, that allow them to be quickly computed 
with arbitrary PDFs, and factorisation and renormalisation scales.
Finally, we show how these formulae can be employed in the construction
of the \amcfast\ bridge that interfaces \applgr\ to \aNLO.

\subsection{The construction of interpolating grids\label{sec:gen}}
The basic idea used by \applgr\ is that of a Lagrange-polynomial
expansion. Given a function $F(z)$ of a real variable $z$, one
has the representation:
\beq
F(z)=\sum_{i=0}^s F\left(\left(\floor{\frac{z}{\delta}-\frac{s-1}{2}}
+i\right)\delta\right)
I_i^{(s)}\left(\frac{z}{\delta}-
\floor{\frac{z}{\delta}-\frac{s-1}{2}}\right)\,,
\label{frepr}
\eeq
where $s$ is a given integer (the interpolation order), 
$\delta$ is a small number (the grid spacing or binning;
$p\delta$ for some integer $p$ is called a grid node),
the interpolating functions are:
\beq
I_i^{(s)}(u)=\frac{(-1)^{s-i}}{i!(s-i)!}
\prod_{k=0,k\ne i}^s (u-k)\,,
\label{Ibasis}
\eeq
and we have denoted by $\floor{u}$ the largest integer smaller
than or equal to $u$:
\beq
\floor{u}\in {\mathbb Z}\,,\;\;\;\;\;\;
u-1<\floor{u}\le u\,,\;\;\;\;\;\;
u\in {\mathbb R}\,.
\eeq
The equality in eq.~(\ref{frepr}) holds up to functional terms
of order $I_i^{(s+1)}$ and higher. Such terms vanish identically
when the argument of $F$ coincides with a grid node, so that 
eq.~(\ref{frepr}) is an identity in that case; this is straightforward
to prove, and follows directly from the values that the interpolating
functions take when computed with an integer argument:
\beq
I_i^{(s)}(k)=\delta_{ik}\,,\;\;\;\;\;\;
0\le k\le s\,,\;\;\;\;\;\;
k\in {\mathbb Z}\,.
\eeq
When $z$ is not a grid node (i.e., $z\ne p\delta$ for any integer $p$),
eq.~(\ref{frepr}) tells one that $F(z)$ is reconstructed by using the 
values that $F$ takes in the $(s+1)$ grid nodes which are nearest to $z$;
the number of relevant nodes to the left of $z$ is equal to number of 
nodes to the right of $z$, possibly up to one.

For any given function $S(z)$, let us now compute the simple 
example integral
\beq
J=\int_a^b dz\, S(z)\, F(z)
\label{Idef}
\eeq
by means of its corresponding Riemann sums (or, equivalently, by
Monte Carlo methods). This implies
\beq
J=\sum_{k=1}^M \Phi_k\, S(z_k)\, F(z_k)\,,
\label{IMC}
\eeq
with $M$ points $z_k\in (a,b)$, and $\Phi_k$ suitable normalisation
factors. By using eq.~(\ref{frepr}), we obtain
\beq
J=\sum_{k=1}^M \Phi_k\, S(z_k)\, 
\sum_{i=0}^s F\Big(\left(p_\delta(z_k)+i\right)\delta\Big)
I_i^{(s)}\left(\frac{z_k}{\delta}-p_\delta(z_k)\right)\,,
\label{I0}
\eeq
where we have defined:
\beq
p_\delta(z)=\floor{\frac{z}{\delta}-\frac{s-1}{2}}\,,
\label{pdeldef}
\eeq
which is the integer associated with the leftmost grid node
in the set of the $(s+1)$ nearest neighbours of $z$. By means of
a change of the summation variable $i$, eq.~(\ref{I0}) becomes:
\beqn
J&=&\sum_{k=1}^M \Phi_k\, S(z_k)\, 
\sum_{j=p_\delta(z_k)}^{s+p_\delta(z_k)} F(j\delta)
I_{j-p_\delta(z_k)}^{(s)}\left(\frac{z_k}{\delta}-p_\delta(z_k)\right)
\label{I1}
\\*&=&
\sum_{j=-\infty}^{\infty} F(j\delta)\, G_j\,,
\label{I2}
\eeqn
with
\beq
G_j=\sum_{k=1}^M \Phi_k\, S(z_k)\, 
I_{j-p_\delta(z_k)}^{(s)}\left(\frac{z_k}{\delta}-p_\delta(z_k)\right)\,
\stepf\left(p_\delta(z_k)\le j\le s+p_\delta(z_k)\right)\,.
\label{griddef}
\eeq
Equation~(\ref{griddef}) defines the grid values $G_j$. Owing to the
$\stepf$ function it contains, the sum in eq.~(\ref{I2}) features a finite
number of non-null contributions (if the range $(a,b)$ is finite).
Thus, the meaning of eq.~(\ref{I2}) is that the integral $J$ can be computed
a posteriori by knowing a finite number of grid values, and the values of 
the function $F$ at the grid nodes; more importantly, this a-posteriori 
computation can be done for any function $F$, because the grid values are 
independent of $F$, and can therefore be pre-evaluated and stored
for a given function $S$. This also explains the reason for writing
the integrand of eq.~(\ref{Idef}) as a product of two functions:
this is convenient whenever the time spent in evaluating $F$ (the
``fast'' function) and $S$ (the ``slow'' function) is small and
large respectively. When this is the case, the computation of the
grid $\{G_j\}$ may be time-consuming, but it is an operation that has 
to be carried out only once; on the other hand, the subsequent evaluations
of eq.~(\ref{I2}) will be quick. We also point out that the derivation
above is unchanged in the case where $z$ is not the integration
variable, but a function itself of one (or more) integration variable(s)
for the problem at hand. This is because the starting point is actually
eq.~(\ref{IMC}), and not eq.~(\ref{Idef}), and in the former the
role of $z_k$ as integration variable can be fully ignored.

\subsection{Generalities on short-distance cross sections\label{sec:obs}}
In what follows we use the expressions derived within the 
FKS subtraction formalism. The same notations as in 
ref.~\cite{Frederix:2011ss} are adopted; this is particularly
convenient in view of the fact that, in that paper, cross sections
were represented in terms of PDF- and scale-independent coefficients,
and these will be the main ingredients in the definition of the
interpolating grids. We point out, however, that the procedure 
outlined below remains valid regardless of the subtraction method chosen 
to perform the NLO computations. Lest we clutter the notation with
details which are irrelevant here, we work by fixing the partonic 
process. This implies, in particular, that we do not need to write 
explicitly the identities of the incoming partons; we shall reinstate 
them later.

The NLO short-distance cross section relevant to a $2\to n+1$ process
consists of four terms:
\beqn
d\sigmaNLO&\longleftrightarrow&\Big\{d\sigmaNLOa\Big\}_{\alpha=E,S,C,SC}\,,
\label{fact3aAPP}
\\
d\sigmaNLOa&=&\fo(\xoa,\muFa)\ft(\xta,\muFa) 
\Wa d\meas_{Bj}d\meas_{n+1}\,,
\label{fact3}
\eeqn
called event ($\alpha=E$), and soft, collinear, and soft-collinear
($\alpha=S, C, SC$) counterevents, respectively. The quantities
$d\meas_{Bj}$ and $d\meas_{n+1}$ are the integration measures over
the Bjorken $x$'s and the $(3n-1)$ phase-space variables respectively,
while $\fo$ and $\ft$ are the PDFs relevant to the colliding partons
coming from the left and from the right. The $\Wa$'s can be parametrised 
as follows:
\beqn
\Wa&=&\gs^{2b+2}(\muRa)\left[
\hWaz+\hWaF\log\!\muFaoQt+\hWaR\log\!\muRaoQt\right]
\nonumber\\*&+&
\gs^{2b}(\muRa)\hWB\delta_{\alpha {\sss S}}\,,
\label{Wadef}
\eeqn
where the coefficients $\widehat{W}$ are (renormalisation and factorisation)
scale- and PDF-independent; the last term on the r.h.s.~of eq.~(\ref{Wadef})
is the Born contribution which, as the notation suggests, factorises $\as^b$.
In eqs.~(\ref{fact3}) and~(\ref{Wadef}) we have denoted by $x_i^{(\alpha)}$,
$\muFa$, and $\muRa$ the Bjorken $x$'s, factorisation scale, and
renormalisation scale respectively; in general, they may assume different
values in the event and various counterevent configurations. Finally,
$Q$ is the Ellis-Sexton scale\footnote{We remind the reader that the 
Ellis-Sexton scale $Q$, originally introduced in ref.~\cite{Ellis:1985er}, 
is any scale that may be used in one-loop computations to express the arguments
of all the logarithms appearing there as \mbox{$s_{ij}/Q^2$} rather
than as \mbox{$s_{ij}/s_{kl}$}, where $s_{ij}$ and $s_{kl}$ are two
invariants constructed with the four-momenta of the particles that
enter the hard process.}. The integration of the above cross section
leads to a set of $4N$ weighted events:
\beqn
\Bigg\{\Big\{\confnpoak\,,\,\xoak\,,\,\xtak\,,\,\wka\Big\}_{\alpha=E,S,C,SC}
\Bigg\}_{k=1}^N\,,
\label{NLOsetApp}
\eeqn
with $\confnpoak$ an $(n+1)$-body kinematic configuration (possibly 
degenerate, in which case it is effectively an $n$-body configuration
that can be used to compute Born matrix elements), and the event weights 
defined as follows:
\beqn
\wka=\frac{d\sigmaNLOa}{d\meas_{Bj}d\meas_{n+1}}
\left(\confnpoak,\xoak,\xtak\right)\,,
\label{wgtNLO}
\eeqn
where we understand possible normalization pre-factors (such as
$1/N$, or adaptive-integration jacobians).
Given an observable $O$, its $h^{th}$ histogram bin defined by
\mbox{$O_{\sss\rm LOW}^{(h)}\le O< O_{\sss\rm UPP}^{(h)}$}
will assume, at the end of the run, the value:
\beqn
\sigma_O^{(h)}&=&
\sum_{k=1}^N\,\sum_\alpha \stepkha\,\wka\,,
\label{histoNLO}
\\
\stepkha&=&
\Theta\!\left(O\left(\confnpoak\right)-O_{\sss\rm LOW}^{(h)}\right)
\Theta\!\left(O_{\sss\rm UPP}^{(h)}-O\left(\confnpoak\right)\right)\,.
\label{binNLO}
\eeqn
In view of eq.~(\ref{Wadef}), it is convenient to recast
eq.~(\ref{histoNLO}) as follows:
\beq
\sigma_O^{(h)}=\sum_{\beta=0,F,R,B}\sigma_{O,\beta}^{(h)}\,,
\label{histoNLO2}
\eeq
where, using eqs.~(\ref{fact3}) and~(\ref{wgtNLO}), one has:
\beqn
\sigma_{O,0}^{(h)}&=&\sum_{k=1}^N\,\sum_\alpha \stepkha\,
\fo(\xoak,\muFak)\,\ft(\xtak,\muFak)\,\gs^{2b+2}(\muRak)\,\hWazk\,,
\label{histoNLOz}
\\
\sigma_{O,F}^{(h)}&=&\sum_{k=1}^N\,\sum_\alpha \stepkha\,
\fo(\xoak,\muFak)\,\ft(\xtak,\muFak)\,\gs^{2b+2}(\muRak)\,
\hWaFk\log\!\muFakoQt\,,
\label{histoNLOF}
\\
\sigma_{O,R}^{(h)}&=&\sum_{k=1}^N\,\sum_\alpha \stepkha\,
\fo(\xoak,\muFak)\,\ft(\xtak,\muFak)\,\gs^{2b+2}(\muRak)\,
\hWaRk\log\!\muRakoQt\,,
\label{histoNLOR}
\\
\sigma_{O,B}^{(h)}&=&\sum_{k=1}^N\,\sum_\alpha \stepkha\,
\fo(\xoak,\muFak)\,\ft(\xtak,\muFak)\,\gs^{2b}(\muRak)\,
\hWBak\,.
\label{histoNLOB}
\eeqn
Here we have defined:
\beqn
\hWabek&=&\hWabe\left(\confnpoak\right)\;\;\;\;\;\;\;\;\beta=0,F,R\,,
\\
\hWBak&=&\hWB\left(\confnpoSk\right)\,\delta_{\alpha {\sss S}}\,,
\\
\muFak&=&\muF\left(\confnpoak\right)\,,
\\
\muRak&=&\muR\left(\confnpoak\right)\,,
\eeqn
namely, the values of the short-distance coefficients and of the scales
computed at the kinematic configurations associated with each of the
events and counterevents of eq.~(\ref{NLOsetApp}).

We now apply the method outlined in sect.~\ref{sec:gen} to 
eqs.~(\ref{histoNLOz})--(\ref{histoNLOB}). In order to do so, 
we recall that the main result of that section is that of computing
the number $J$, whose original expression is that of eq.~(\ref{IMC})
(regardless of the fact that such an expression was in turn obtained
from the integral of eq.~(\ref{Idef}), for the reason explained at
the end of sect.~\ref{sec:gen}), by means of the interpolating grid
given in eq.~(\ref{griddef}) and of eq.~(\ref{I2}).

One starts by observing that the $h^{th}$ bin value $\sigma_{O,\beta}^{(h)}$ 
of eqs.~(\ref{histoNLOz})--(\ref{histoNLOB}) is written in the same form as 
$J$ of eq.~(\ref{IMC}): the double sums over $k$ and $\alpha$
in eqs.~(\ref{histoNLOz})--(\ref{histoNLOB}) can easily be recast in 
the form of a single sum as in eq.~(\ref{IMC}). The goal is therefore that
of representing $\sigma_{O,\beta}^{(h)}$ with an interpolating grid, as
is done for $J$ in eq.~(\ref{I2}). In order to do so, one may easily
proceed by analogy. First of all, since we are interested in
a quick recomputation of the cross section after changing the scales
and the PDFs, and given that the latter depend on the Bjorken $x$'s
and the factorisation scale, the implication is that the role of the 
variable $z$ of eqs.~(\ref{IMC})--(\ref{griddef}) needs to be
played by the four variables:
\beq
\xo,\;\xt,\;\muF,\;\muR\,.
\label{varset}
\eeq
In other words, we have the correspondence:
\beq
\{z_k\}_{k=1,M}\;\;\leftrightarrow\;\;
\Big\{\xoak,\xtak,\muFak,\muRak\Big\}_{k=1,N}^{\alpha=E,S,C,SC}\,.
\eeq
The polynomial expansion of eq.~(\ref{frepr}), whence the interpolating
grid is derived, is valid for each of the variables in eq.~(\ref{varset});
therefore, the only change w.r.t.~the case of the computation of $J$ is 
the fact that the one-dimensional grid $\{G_j\}$ defined in 
eq.~(\ref{griddef}) will be replaced by a four-dimensional grid,
corresponding to the variables in eq.~(\ref{varset}).
For the construction of the latter, it is sufficient to compare
directly eq.~(\ref{IMC}) with eqs.~(\ref{histoNLOz})--(\ref{histoNLOB}).
The role of the ``slow'' and ``fast'' functions $S$ and $F$ will be
played by the short-distance coefficients $\widehat{W}$ and by the
scale- and PDF-dependent terms respectively; that of the factor
$\Phi_k$ by the bin-defining $\stepkha$ of eq.~(\ref{binNLO}).
In other words, we have the following identifications:
\beqn
\beta=0\;\;\Longrightarrow\;\;\;\;
(S,F)&\leftrightarrow&
\left(\hWz\,,\;\fo(\xo,\muF)\ft(\xt,\muF)\gs^{2b+2}(\muR)\right)\,,
\label{NLOgr1}
\\
\beta=F\;\;\Longrightarrow\;\;\;\;
(S,F)&\leftrightarrow&
\left(\hWF\,,\;\fo(\xo,\muF)\ft(\xt,\muF)\gs^{2b+2}(\muR)\log\!\muFoQt\right)\,,
\label{NLOgr2}
\\
\beta=R\;\;\Longrightarrow\;\;\;\;
(S,F)&\leftrightarrow&
\left(\hWR\,,\;\fo(\xo,\muF)\ft(\xt,\muF)\gs^{2b+2}(\muR)\log\!\muRoQt\right)\,,
\label{NLOgr3}
\\
\beta=B\;\;\Longrightarrow\;\;\;\;
(S,F)&\leftrightarrow&
\left(\hWB\,,\;\fo(\xo,\muF)\ft(\xt,\muF)\gs^{2b}(\muR)\right)\,,
\label{NLOgr4}
\eeqn
and
\beq
\Phi_k\;\leftrightarrow\;\stepkha\,.
\label{NLOgr5}
\eeq
It is now sufficient to use eqs.~(\ref{NLOgr1})--(\ref{NLOgr5}) to
rewrite eqs.~(\ref{I2}) and~(\ref{griddef}). We denote by $j_1$, $j_2$,
$j_3$, and $j_4$ the indices that run over the grid nodes relevant
to the variables $\xo$, $\xt$, $\muF$, and $\muR$, respectively;
$\delta_i$ and $s_i$, $i=1,\ldots 4$, are the corresponding grid spacings
and interpolating orders. We obtain:
\beqn
\sigma_{O,0}^{(h)}&=&\sum_{j_1,j_2,j_3,j_4}
\fo(j_1\delta_1,j_3\delta_3)\,\ft(j_2\delta_2,j_3\delta_3)\,
\gs^{2b+2}(j_4\delta_4)\,G_{j_1j_2j_3j_4}^{(h,0)}\,,
\label{gridNLOz}
\\
\sigma_{O,F}^{(h)}&=&\sum_{j_1,j_2,j_3,j_4}
\fo(j_1\delta_1,j_3\delta_3)\,\ft(j_2\delta_2,j_3\delta_3)\,
\gs^{2b+2}(j_4\delta_4)
\log\left(\!\frac{j_3\delta_3}{\QESa}\!\right)^2
G_{j_1j_2j_3j_4}^{(h,F)}\,,
\label{gridNLOF}
\\
\sigma_{O,R}^{(h)}&=&\sum_{j_1,j_2,j_3,j_4}
\fo(j_1\delta_1,j_3\delta_3)\,\ft(j_2\delta_2,j_3\delta_3)\,
\gs^{2b+2}(j_4\delta_4)
\log\left(\!\frac{j_4\delta_4}{\QESa}\!\right)^2
G_{j_1j_2j_3j_4}^{(h,R)}\,,
\label{gridNLOR}
\\
\sigma_{O,B}^{(h)}&=&\sum_{j_1,j_2,j_3,j_4}
\fo(j_1\delta_1,j_3\delta_3)\,\ft(j_2\delta_2,j_3\delta_3)\,
\gs^{2b}(j_4\delta_4)\,G_{j_1j_2j_3j_4}^{(h,B)}\,,
\label{gridNLOB}
\eeqn
with the interpolating grids:
\beqn
G_{j_1j_2j_3j_4}^{(h,\beta)}&=&
\sum_{k=1}^N\,\sum_\alpha \stepkha\,\hWabek\, 
\label{4dimgrid}
\\&&\phantom{aaa}\times
I_{j_1-p_{\delta_1}(\xoak)}^{(s_1)}\!
\left(\frac{\xoak}{\delta_1}-p_{\delta_1}(\xoak)\right)
\stepf\!\left(p_{\delta_1}(\xoak)\le j_1\le s_1+p_{\delta_1}(\xoak)\right)
\nonumber
\\&&\phantom{aaa}\times
I_{j_2-p_{\delta_2}(\xtak)}^{(s_2)}\!
\left(\frac{\xtak}{\delta_2}-p_{\delta_2}(\xtak)\right)
\stepf\!\left(p_{\delta_2}(\xtak)\le j_2\le s_2+p_{\delta_2}(\xtak)\right)
\nonumber
\\&&\phantom{aaa}\times
I_{j_3-p_{\delta_3}(\muFak)}^{(s_3)}\!
\left(\frac{\muFak}{\delta_3}-p_{\delta_3}(\muFak)\right)
\stepf\!\left(p_{\delta_3}(\muFak)\le j_3\le s_3+p_{\delta_3}(\muFak)\right)
\nonumber
\\&&\phantom{aaa}\times
I_{j_4-p_{\delta_4}(\muRak)}^{(s_4)}\!
\left(\frac{\muRak}{\delta_4}-p_{\delta_4}(\muRak)\right)
\stepf\!\left(p_{\delta_4}(\muRak)\le j_4\le s_4+p_{\delta_4}(\muRak)\right)\,.
\nonumber
\eeqn
Equations~(\ref{gridNLOz})--(\ref{4dimgrid}) give the most general
representation of NLO cross sections in terms of interpolating grids.
We shall employ them (in a simplified form) in the next subsection,
in the construction of the \amcfast\ bridge to \applgr.

\subsection{The interface of {\sc APPLgrid} with 
{\sc MadGraph5\_aMC@NLO}\label{sec:int}}

\applgr\ is a
general-purpose {\sc C++} library that provides a suitable number of
interpolation and convolution routines that can be used to construct fast
interfaces to NLO calculations of lepton-proton and hadron-hadron
collider processes. \applgr\ is widely used by various PDF fitting 
collaborations as well as by ATLAS and CMS in their own PDF studies.
Our purpose is that of exploiting \applgr\ for the construction
of the interpolating grids of eq.~(\ref{4dimgrid}), and their subsequent
use in the calculation of the histogram bins, 
eqs.~(\ref{gridNLOz})--(\ref{gridNLOB}). As those equations show,
\applgr\ needs to be given, in an initialisation phase, the observables
to be computed and their respective binnings (we stress again that each
bin of each observable corresponds to a set of four grids per type of
parton luminosity); and, on an event-by-event basis, the values of those
observables and the short-distance coefficients $\widehat{W}$.
These tasks are essentially what \amcfast\ is responsible for:
it extracts the relevant information from \aNLO, and feeds them
to \applgr, in the format required by the latter.

The first observation is that, in its current version, \applgr\ supports
three-dimensional grids, while those in eq.~(\ref{4dimgrid}) are
four-dimensional. The reason for the latter is that in our derivation
we have left the freedom of choosing different functional forms for
the factorisation and renormalisation scales. On the other hand,
such a flexibility is seldom exploited, and typically
one chooses $\muF$ and $\muR$ equal in the whole phase space,
up to an overall constant. This is equivalent to setting:
\beq
\muF=\xiF\,\mu\,,\;\;\;\;\;\;\;\;
\muR=\xiR\,\mu\,,
\label{FeqR}
\eeq
with $\mu$ a function of the kinematics.
When doing this, two of the four variables in eq.~(\ref{varset}) 
are degenerate, and therefore one must simply consider:
\beq
\xo,\;\xt,\;\mu\,,
\label{varset2}
\eeq
which reduces the number of grid dimensions from four to three.
In this case, it is also convenient to set\footnote{Although we did not
indicate this explicitly in sect.~\ref{sec:obs}, the Ellis-Sexton
scale is in general a function of the kinematics, whence the possibility
of using eq.~(\ref{Qeqmu}).}:
\beq
Q=\mu\,,
\label{Qeqmu}
\eeq
so that 
\beq
\log\left(\!\frac{j_3\delta_3}{\QESa}\!\right)^2\,\longrightarrow\,
\log\xiF^2\,,\;\;\;\;\;\;\;\;
\log\left(\!\frac{j_4\delta_4}{\QESa}\!\right)^2\,\longrightarrow\,
\log\xiR^2\,,
\eeq
in eqs.~(\ref{gridNLOF}) and~(\ref{gridNLOR}) respectively.
The other changes to eqs.~(\ref{gridNLOz})--(\ref{4dimgrid}) due to
the reduction of the grid dimensionality are all trivial: one eliminates
the sums over $j_4$, formally replaces:
\beq
j_3\delta_3\,\longrightarrow\,j_3\delta_3\,\xiF\,,\;\;\;\;\;\;\;\;
j_4\delta_4\,\longrightarrow\,j_3\delta_3\,\xiR\,,
\label{repl}
\eeq
and $\muF$ with $\mu$ in the next-to-last line on the r.h.s.~of 
eq.~(\ref{4dimgrid}) (this and eq.~(\ref{repl}) are due to the fact that 
now it is the variable $\mu$ which corresponds to one of the dimensions 
of the grid), and finally eliminates the last line of that equation.

The next thing to consider is that \applgr\ does not use the variables
in eq.~(\ref{varset2}) directly, but rather constructs the grids using:
\beq
\yo,\;\yt,\;\tau\,,
\label{varset3}
\eeq
which are defined through a change of variables, several of which 
are available but is usually set to:
\beqn
&&y_i=Y(x_i)\;\;\;\;\;\;\;\;
Y(x)=\log\frac{1}{x}+\kappa (1-x)\,,
\label{ydef}
\\
&&\tau=T(\mu)=\log\log\frac{\mu^2}{\Lambda^2}\,.
\label{taudef}
\eeqn
This is because the grids feature a linear spacing, and given
the PDFs and $\as$ dependences on Bjorken $x$'s and scales a linear
sampling in terms of the variables of eq.~(\ref{varset3}) turns out to
be more effective than one based the variables of eq.~(\ref{varset2}).
In eq.~(\ref{ydef}) the parameter $\kappa$ controls the relative density
of grid nodes in the large- w.r.t.~the small-$x$ region; in 
eq.~(\ref{taudef}), the parameter $\Lambda$ is typically chosen
to be of the order of $\Lambda_{\rm\sss QCD}$.

Finally, one needs to consider the fact that the formulae presented
above are obtained for a given partonic process. In order to compute
the actual hadronic cross section, one needs to sum over all possible
such processes, so that eq.~(\ref{fact3}) must be generalised as 
follows:
\beq
d\sigmaNLOa=\sum_{rsq}\fr(\xoa,\muFa)\fs(\xta,\muFa) 
\Warsq d\meas_{Bj}d\meas_{n+1}\,,
\label{fact4}
\eeq
where $r$ and $s$ are the identities of the incoming partons, and
$q$ collectively denotes the identities of all outgoing partons. Note 
that, for simple-enough cases, the sum over $q$ is trivial, since $(r,s)$ 
are sufficient to determine unambiguously a partonic process; however,
the notation used in eq.~(\ref{fact4}) is general, and encompasses
all possible situations.
Since the interpolating grids are additive, the most straightforward
solution is that of following the procedure outlined so far, and of 
creating a grid for each possible partonic process. There is however
a better strategy, that helps save disk space and decrease memory footprint, 
in that it reduces
the number of interpolating grids. This is based on the observation
that one may find pairs of parton indices $(r,s)$ and $(r^\prime,s^\prime)$
such that:
\beq
\Warsq=\Warsqp\;\;\;\;\;\;{\rm with}\;\;\;\;\;\;
(r,s)\ne (r^\prime,s^\prime)
\;\;\;{\rm for~some}\;\;q\,,\;q^\prime\,.
\label{WeqW}
\eeq
This suggests to rewrite eq.~(\ref{fact4}) in the following way:
\beq
d\sigmaNLOa=\sum_{l}\luml(\xoa,\xta,\muFa) 
\sum_{q}\Walq d\meas_{Bj}d\meas_{n+1}\,,
\label{fact5}
\eeq
where
\beq
\luml(\xoa,\xta,\muFa)=\sum_{rs}T_{rs}^{(l)}\fr(\xoa,\muFa)\fs(\xta,\muFa)\,,
\label{lumldef}
\eeq
and the values of $T_{rs}^{(l)}$ are either zero or one; we have
implicitly defined $\Walq=\Warsq$ for $(r,s,q)$ and $l$ such that 
$T_{rs}^{(l)}=1$.
We point out that, while there may be more than one way to write the
r.h.s.~of eq.~(\ref{fact5}) (in other words, the luminosity factors
$\luml$ may not be uniquely defined), it is always possible to arrive
at such a form, thanks to the fact that the r.h.s.'s of eqs.~(\ref{fact4})
and~(\ref{fact5}) are strictly identical. In fact, eq.~(\ref{fact5}) 
is what is used internally by \aNLO; one of the tasks of \amcfast\ is
that of gathering this piece of information, and of using it to construct 
the luminosity factors $\luml$ to be employed at a later stage.
In eq.~(\ref{fact5}) one factors out identical matrix elements; since the
computation of these is the most time-consuming operation, this procedure
helps increase the overall efficiency, which is larger the larger the number
of terms in each luminosity factor $\luml$. The fact that, in general,
the number of terms in the sum over $l$ in eq.~(\ref{fact5}) is smaller
than that in the sums over $(r,s)$ in eq.~(\ref{fact4}) is ultimately
what allows one to reduce the number of interpolating grids.
The counting of the terms in such sums is easy when $(r,s)$
are sufficient to determine uniquely the partonic process.
Denoting by $n_l$ the largest value assumed by the luminosity 
index $l$ for a given process:
\beq
1\le l\le n_l\,,
\eeq
and by $\NF$ the number of active light flavours, one has:
\beq
1\le n_l\le (2\NF+1)^2\,,
\eeq
with the two limiting cases being either that where all of the allowed PDF
combinations are associated with the same partonic matrix element, or that
where each PDF combination corresponds to a different partonic matrix element.
In appendix~\ref{sec:lum} we shall give the explicit
form of eq.~(\ref{lumldef}) for all of the processes studied.

By putting everything together, we finally arrive at the representation
of the $h^{th}$ bin value and of its corresponding grids as they are
constructed by the \aNLO\ -- \amcfast\ -- \applgr\ chain.
We denote by $\delta_y$ and $s_y$ the grid spacing
and interpolating order relevant to the variables $y_1$ and $y_2$;
the analogous quantities relevant to the variable $\tau$ are
denoted by $\delta_\tau$ and $s_\tau$ respectively.
We have:
\beqn
\sigma_{O,0}^{(h)}&=&\sum_{j_1,j_2,j_3}\sum_{l=1}^{n_l}
\hluml(j_1\delta_y,j_2\delta_y,j_3\delta_\tau)
\hgs^{2b+2}(j_3\delta_\tau)\,G_{j_1j_2j_3}^{(h,0,l)}\,,
\label{appNLOz}
\\
\sigma_{O,F}^{(h)}&=&\sum_{j_1,j_2,j_3}\sum_{l=1}^{n_l}
\hluml(j_1\delta_y,j_2\delta_y,j_3\delta_\tau)\,
\hgs^{2b+2}(j_3\delta_\tau)\,
\log\xiF^2\,G_{j_1j_2j_3}^{(h,F,l)}\,,
\label{appNLOF}
\\
\sigma_{O,R}^{(h)}&=&\sum_{j_1,j_2,j_3}\sum_{l=1}^{n_l}
\hluml(j_1\delta_y,j_2\delta_y,j_3\delta_\tau)\,
\hgs^{2b+2}(j_3\delta_\tau)\,
\log\xiR^2\,G_{j_1j_2j_3}^{(h,R,l)}\,,
\label{appNLOR}
\\
\sigma_{O,B}^{(h)}&=&\sum_{j_1,j_2,j_3}\sum_{l=1}^{n_l}
\hluml(j_1\delta_y,j_2\delta_y,j_3\delta_\tau)\,
\hgs^{2b}(j_3\delta_\tau)\,
G_{j_1j_2j_3}^{(h,B,l)}\,,
\label{appNLOB}
\eeqn
with the grids:
\beqn
G_{j_1j_2j_3}^{(h,\beta,l)}&=&
\sum_{k=1}^N\,\sum_\alpha\,\sum_q \stepkha\,\hWabelqk\, 
\label{3dimgrid}
\\&&\phantom{aaa}\times
I_{j_1-p_{\delta_y}(\yoak)}^{(s_y)}\!
\left(\frac{\yoak}{\delta_y}-p_{\delta_y}(\yoak)\right)
\stepf\!\left(p_{\delta_y}(\yoak)\le j_1\le s_y+p_{\delta_y}(\yoak)\right)
\nonumber
\\&&\phantom{aaa}\times
I_{j_2-p_{\delta_y}(\ytak)}^{(s_y)}\!
\left(\frac{\ytak}{\delta_y}-p_{\delta_y}(\ytak)\right)
\stepf\!\left(p_{\delta_y}(\ytak)\le j_2\le s_y+p_{\delta_y}(\ytak)\right)
\nonumber
\\&&\phantom{aaa}\times
I_{j_3-p_{\delta_\tau}(\tauak)}^{(s_\tau)}\!
\left(\frac{\tauak}{\delta_\tau}-p_{\delta_\tau}(\tauak)\right)
\stepf\!\left(p_{\delta_\tau}(\tauak)\le j_3\le s_\tau+p_{\delta_\tau}
(\tauak)\right)\,.
\nonumber
\eeqn
In eq.~(\ref{3dimgrid}) we have inserted the partonic indices 
$l$ and $q$ in the short-distance quantities $\widehat{W}$, and we 
have used eqs.~(\ref{ydef}) and~(\ref{taudef}) to introduce:
\beqn
\yiak&=&Y(\xiak)\,,
\\
\tauak&=&T\left(\mu\left(\confnpoak\right)\right)\,,
\eeqn
and we have defined (note the factors $\xiF$ and $\xiR$):
\beqn
\hluml(\yo,\yt,\tau)&=&
\luml\left(Y^{-1}(\yo),Y^{-1}(\yt),\xiF\, T^{-1}(\tau)\right)\,,
\\
\hgs(\tau)&=&\gs\left(\xiR\, T^{-1}(\tau)\right)\,.
\eeqn

Equations~(\ref{appNLOz})--(\ref{3dimgrid}) are the main results
of this paper. In the initialisation phase, \amcfast\ provides
\applgr\ with the total number of grids needed (equal to the sum 
over all observables of the number of bins relevant to each observable,
times four), the grid spacings $\delta_y$ and $\delta_\tau$,
the interpolation orders $s_y$ and $s_\tau$, and the interpolation
ranges in $y_i$ and $\tau$. These information are under the user's
control; in particular, one must make sure that the latter ranges 
are sufficiently wide for the process under consideration, and one will
want to be careful in the case of a dynamical scale choice for $\mu$.
Then, during the course of the run and event-by-event, \amcfast\ gets
$\stepkha$, $\hWabelqk$, $\yiak$, and $\tauak$ from \aNLO\ and
feeds\footnote{\amcfast-specific input routines have been added
to \applgr.} them to \applgr, whose 
grid-filling internal routines iteratively construct 
$G_{j_1j_2j_3}^{(h,\beta,l)}$ as defined in eq.~(\ref{3dimgrid}).

We conclude this section with two remarks. Firstly, the sums over
$l$ and $q$ in the formulae above achieve the sum over subprocesses
which is necessary in order to obtain the hadronic cross section.
In practice, in \aNLO\ the number of contributions which are integrated
separately and eventually summed to give the physical cross section is
often larger than the number of subprocesses, owing to the FKS dynamic
partition and to the use of multi-channel techniques (see 
ref.~\cite{Alwall:2014hca} for more details). When interfacing
\aNLO\ with \applgr, each of these contributions generates temporary
grids, which are then combined by \amcfast\ at the end of
the run to give the actual grids of 
eq.~(\ref{3dimgrid}) that are to be used in fast computations
(i.e., in eqs.~(\ref{appNLOz})--(\ref{appNLOB})). Secondly, thanks to the 
fact that the information on the cross section is (also) given in terms
of the scale- and PDF-independent coefficients $\widehat{W}$,
\aNLO\ is capable of computing a-posteriori PDF and scale
uncertainties independently of its interface to \applgr, by means of
a reweighting procedure
(see ref.~\cite{Frederix:2011ss}). While this feature renders the
recomputation of the cross section with alternative PDFs and/or scales
much faster than the original calculation, it is still not fast enough
to be employed in PDF fits, for which the only viable solution
is that of using the interpolating grids discussed here. The reason is
related to the sum over events ($1\le k\le N$ and $\alpha=E,S,C,SC$
in the previous formulae): while such sum is performed only once
in the case of interpolating grids (when the grid is constructed:
see eq.~(\ref{3dimgrid})), it must be carried out for each new choice
of PDFs and scales in the context of reweighting. This difference is
crucial, because $N$ must be a large number (especially so in fNLO
computations), in order to obtain a good statistical 
precision\footnote{On the other hand, reweighting is 
more flexible than grid interpolation, in that it gives one the possibility
of recomputing the cross section by adopting a different {\em functional}
form for the scales w.r.t.~that used in the original computation, which
is not feasible when using grids. Such a possibility is however not available
in the public version of \aNLO. More importantly, reweighting does not
need the prior knowledge of the observables which one is going to study.}.

\subsection{Scale choices\label{sec:scale}}
As one can see from eq.~(\ref{Wadef}), the grids 
$G_{j_1j_2j_3}^{(h,0,l)}$ and $G_{j_1j_2j_3}^{(h,B,l)}$ are 
all that is needed when one is interested in computing the cross section
that corresponds to setting the factorisation and renormalisation scales 
equal to their reference value $\mu$, i.e.~with $\xiF=\xiR=1$
(see eq.~(\ref{FeqR})). When $\xiF\ne 1$ and/or $\xiR\ne 1$, then
the grids $G_{j_1j_2j_3}^{(h,F,l)}$ and/or $G_{j_1j_2j_3}^{(h,R,l)}$,
respectively, are also necessary (see eqs.~(\ref{appNLOF})
and~(\ref{appNLOR})). We point out that these two grids
are not constructed when \applgr\ is interfaced to codes other 
than \aNLO. In that case, \applgr\ is still capable of computing
the cross section for non-central scales, since the form of the latter can be 
determined by using renormalisation group equations (RGEs). By doing so,
one arrives at an expression (e.g., eq.~(14) of ref.~\cite{Carli:2010rw}) 
which is in one-to-one correspondence with the results derived here (in 
order to see this, one must use the explicit expressions of the $\widehat{W}$ 
coefficients given in ref.~\cite{Frederix:2011ss}); this is not
surprising, since both are ultimately a direct consequence of 
RGE invariance. 

Although fully equivalent from a physics viewpoint,
the two-grid and four-grid approaches do not give pointwise-identical 
results (in other words, their outcomes are strictly equal only in
the limit of infinite statistics). The main advantage of using 
only two grids is that of a smaller memory footprint.
Conversely, the two-grid approach has two drawbacks that
we consider significant, and are the reason why \amcfast\ works
with four grids. Firstly, the $\xiF$-dependent term features a 
convolution (i.e., a one-dimensional integral) of a one-loop 
Altarelli-Parisi kernel~\cite{Altarelli:1977zs} with a PDF, 
$P^{(0)}\otimes f$. This convolution is effectively performed 
when summing over events in the four-grid approach, in 
eq.~(\ref{appNLOF}). The absence of $G_{j_1j_2j_3}^{(h,F,l)}$
in the two-grid procedure implies that \applgr\ needs to perform
this convolution explicitly, and it does so by resorting to an
{\em external} code, the PDF-evolution package 
{\sc\small HOPPET}~\cite{Salam:2008qg}. \applgr\ does work 
without {\sc\small HOPPET} being installed, but in this
case it is not possible to choose a non-central factorisation scale.
Secondly, and related to the previous point: the representation
of a cross section in terms of interpolating grids is exact up
to terms of higher order in the interpolation-order parameter;
this implies, for example, that eq.~(\ref{appNLOF}) is an identity
up to terms that contain the interpolating functions $I^{(s_y+1)}$
and $I^{(s_\tau+1)}$ -- as we shall show later, in all practical
cases such missing terms are totally negligible. The crucial point here
is that this is an identity regardless of the statistics used
in the computation of the cross section (i.e., it is independent of $N$, 
up to fluctuations that may affect the grid construction in the case 
of very small $N$'s; again, we shall document this later). However, and 
for the specific case of the $\xiF$-dependent eq.~(\ref{appNLOF}), this is
true only because the convolution integral $P^{(0)}\otimes f$
implicit in there is carried out by the very procedure (i.e., the
same number of events, the same seeds) that fills the interpolating
grid. Any other way of computing $P^{(0)}\otimes f$, and in particular
one that uses an external program such as {\sc\small HOPPET}, implies
that this property does not hold. Therefore, in the two-grid approach
the difference between a cross section computed with non-central scales,
and its representation in term of grids, is much larger than formally
guaranteed by the chosen interpolation orders; this difference
tends to zero only in the limit of infinite statistics $N\to\infty$.
We did explicitly verify that this is indeed the case.

In summary, we believe that working with four grids in \applgr\
gives a couple of clear advantages over the two-grid procedure:
one does not need to install any external convolution package,
and the cross section and its grid representation are identical
for any scale choices, regardless of the statistics used. The 
latter feature is beneficial also in view of the fact that results
for central and non-central scale choices in \aNLO\ are fully 
correlated: hence, the ratios of predictions obtained with different
scales are more stable than each of them individually.
We point out that both of these advantages are relevant to the
factorisation scale dependence. As far as the renormalisation
scale is concerned, the situation is simpler, since no convolution
integral is involved. Therefore, the optimal approach\footnote{This is
only the case if one does not need to deal with processes that feature
$\muR$-dependent Yukawa; otherwise, the fourth grid is not trivially
related to the Born one.} would be
that of using three grids, and construct $G_{j_1j_2j_3}^{(h,R,l)}$
(relevant to the $\xiR$ dependence) on the fly, using
$G_{j_1j_2j_3}^{(h,B,l)}$. However, the gain of doing so w.r.t.~our 
four-grid implementation is extremely marginal, and thus we did not 
consider it when constructing \amcfast.

\section{Results and validation\label{sec:pheno}}
As an illustration of the flexibility of \amcfast, in this section we 
present predictions for a variety of LHC processes, which either are 
currently or might soon become relevant for PDF determinations.
Predictions are given for a 14~TeV LHC, using as inputs the 
NNPDF2.3 NLO PDF set~\cite{Ball:2012cx} (associated with $\as(M_Z)=0.1180$)
in the case of five or four light flavours, and the NNPDF2.1 
NLO PDF set~\cite{Ball:2011mu} (associated with $\as(M_Z)=0.1190$)
in the case of three light flavours.
We have made this choice of input parameters in order to be
definite: the pattern of our findings would be unchanged had we
used e.g.~different PDF sets~\cite{Ball:2012wy,Martin:2009iq,Gao:2013xoa}.
For the same reason, we do not consider PDF uncertainties in our 
illustrative study, and thus we only employ the central PDF member
of the NNPDF2.3 and NNPDF2.1 sets. All such sets are taken
from the {\sc\small LHAPDF5}~\cite{Bourilkov:2006cj} library.

\subsection{General strategy}
The main idea is the following: given a process, an observable $O$,
and a binning for the differential distribution in $O$, we compute the
value of its $h^{th}$ bin (for all bins) $\sigma_O^{(h)}$,
in two different ways: directly, by means of \aNLO, and a posteriori,
using the grids constructed with the \aNLO\ -- \amcfast\ -- \applgr\ 
chain. For the latter chain to be validated, these two results must 
be identical up to numerical inaccuracies; we shall call the former
the {\em reference} result, and the latter the {\em reconstructed}
result. In other words, we regard
the l.h.s.~of eq.~(\ref{histoNLO2}) as computed directly with
\aNLO, and its r.h.s.~as obtained from the interpolating grids,
eqs.~(\ref{appNLOz})--(\ref{3dimgrid}), so as to establish 
numerically the accuracy of the equality in eq.~(\ref{histoNLO2}).
As we have already stressed, we expect that inaccuracies are solely
due the interpolating procedure, and not to a (possible) lack of
statistics in the simulations; importantly, as was discussed in 
sect.~\ref{sec:scale}, in our approach this property must hold for 
any non-central scale choices as well. In order to document this,
all of our results will be presented for three different scale 
choices\footnote{One should be careful not to interpret the envelope 
resulting from eqs.~(\ref{sc1})--(\ref{sc3}) as representative of
the higher-order theoretical uncertainty: these choices are 
made for the sole purpose of validating \amcfast\ with arbitrary
$\xiF$ and $\xiR$.}:
\beqn
\muF=\mu\,,&&\muR=\mu\,,
\label{sc1}
\\
\muF=2\mu\,,&&\muR=\mu/2\,,
\label{sc2}
\\
\muF=\mu/2\,,&&\muR=2\mu\,,
\label{sc3}
\eeqn
and as obtained from two different runs: one with ``low'' statistics,
and one with ``high'' statistics. For each choice of scales and type
of run, we shall plot the reference result, and the ratio of the
reconstructed over the reference result. From that ratio, we shall
see that the typical numerical inaccuracies due to the use of the
interpolating grids are in the lower $10^{-4}$, with occasional larger
differences in the case of the low-statistics runs, due to a 
still-insufficient amount of information stored in the 
grids\footnote{This inaccuracy includes a contribution from the
internal interpolation in {\sc\small LHAPDF5},
which is significantly reduced with {\sc\small LHAPDF6}.}.
The crucial point is the following: even with very limited statistics,
the reconstruction of the reference results is extremely good, and this
for distributions whose quality is largely insufficient for any
phenomenological application. This proves that, for all practical
purposes, the level of agreement between reconstructed and reference 
results is independent of the statistics employed.

In keeping with the strategy at the core of all applications of \aNLO,
the only operations to be performed by the user when running with
\amcfast\ are related to the process-specific analysis. In particular,
in view of the fast interface to fNLO results, the user's analysis
(to be stored, as for any other kind of fNLO user utility, into the
directory {\tt MYPROC/FixedOrderAnalysis} -- see sect.~3.2.2 of
ref.~\cite{Alwall:2014hca} for more details) will need to
contain, for each observable that he/she wants to consider, 
the instructions for filling the histogram that will contain
the reference result (such instructions are identical to those of
an fNLO analysis used when no interface with \amcfast\ is 
considered), and those for setting up the associated
interpolating grids. Several templates
of analyses that can be used with \amcfast\ will be provided in
{\tt MYPROC/FixedOrderAnalysis} (starting from the next version
of \aNLO), and the user should be able to easily adapt them to suit the
needs of any given calculation. For the whole procedure to work,
one will need to install, on top of \aNLO\ and of the bridge code
\amcfast, also \applgr\ version 1.4.56 or higher.

\begin{table}[h]
\begin{center}
\begin{tabular}{cc|cc}
\toprule
Parameter & value & Parameter & value
\\\midrule
$\kappa$ & 5 & $\Lambda$ & 0.250~GeV\\
$(x_{\min},x_{\max})$ & $(2\mydot 10^{-7},1)$ &
$(\mu_{\min},\mu_{\max})$ & (10,3162)~GeV \\
$N_y$ & 50 &
$N_\tau$ & 30 \\
$s_y$ & 3 &
$s_\tau$ & 3 \\
\bottomrule
\end{tabular}
\end{center}
\caption{\label{tab:param}
Grid parameters used in the \aNLO\ -- \amcfast\ -- \applgr\ validation
runs presented here. In the case of $t\bt$ production (sect.~\ref{sec:tt}),
the values of $\mu_{\max}$ and $N_\tau$ have been set equal to $10^4$~GeV and
$50$, respectively.
}
\end{table}
All of the calculations reported below are carried out either with 
the default physics model of fNLO computations in \aNLO, which 
corresponds to the SM with $\NF=4$ light flavours, or with 
its $\NF=5$ or $\NF=3$ variants; the quarks which are not light 
(the top and, depending on the value of $\NF$, the bottom and the 
charm) are given a non-zero mass. The parameters used to initialise 
the interpolating grids are given in table~\ref{tab:param}:
from the first and second lines there, and by using eqs.~(\ref{ydef})
and~(\ref{taudef}), one immediately obtains the actual interpolation
ranges in the variables $y_i$ and $\tau$, and the corresponding
grid spacings by taking the ratio of those over $(N_y-1)$ and
$(N_\tau-1)$, respectively. In the low- and high-statistics runs
we employ a number of phase-space points per integration channel
of the order of $10^3$ and $10^6$ respectively. For each
process we shall consider a couple of observables to be definite,
but we emphasise again that the number and the nature of the 
observables that one can compute are the user's choice.
The  kinematical cuts will in general be similar to those
used by ATLAS or CMS in some analogous experimental analyses;
however, a comparison with experimental data is beyond 
the scope of this work. In all cases, we shall set $\mu=H_{\sss T}/2$,
with $H_{\sss T}$ the scalar sum of the transverse energies of all
the final-state particles. We point out that the choice of scales
may have some implications on the settings of the corresponding
interpolation parameters. For example, as one can see from 
table~\ref{tab:param}, in the case of $t\bt$ production the values
of $\mu_{\max}$ and $N_\tau$ have been increased w.r.t.~those used
for all the other processes, and this in order to attain a comparable
numerical precision. On the other hand, we have verified that, by choosing
scales which do not depend on the event kinematics, and are of the
order of the hardness of the given process at threshold, setting
$N_\tau=10$ is amply sufficient. To put things in perspective, however,
we emphasise that even by using $N_\tau=30$ in $t\bt$ production with
$\mu=H_{\sss T}/2$, numerical inaccuracies are at most $5\mydot 10^{-4}$ 
and generally smaller than that, i.e.~better than
any PDF-fit application will ever need.

We start by studying top-quark pair production, which provides
information on the gluon PDF in the large-$x$ region complementary 
to those provided by jet-production data. We shall then consider
isolated-photon production in association with jets, a process that allows a
variety of tests of perturbative QCD and that is also directly sensitive to
the gluon PDF. We shall next discuss the production of a lepton pair in
association with one extra hard jet. We shall then turn to 
considering $Zb\bar{b}$ production, a case-study of how the complete 
automation of fast NLO QCD calculations allows one to include in global
PDF fits arbitrarily complicated processes at no significant extra cost.
Finally, we shall discuss in some details another important process
at the LHC for PDF extraction, namely the production of a $W$ boson in 
association with charm quarks, which is directly sensitive to the 
poorly known strange-quark PDF, and for which data from ATLAS and CMS 
have recently become available.

\subsection{Top-quark pair production\label{sec:tt}}
Top-quark pair production provides one with unique information on the
poorly known large-$x$ gluon PDF~\cite{Czakon:2013tha,Alekhin:2013nda,
Guzzi:2014wia}.
Unlike the case of the Tevatron, where $t\bt$ production is mostly driven
by $q\bar{q}$ scattering, at the LHC the $gg$-initiated process is dominant,
contributing to more that 85\% of the total inclusive
cross section. The recent calculation of the total
cross section to NNLO~\cite{Czakon:2013goa}, as well as the upcoming NNLO
results for differential distributions, make top-quark pair production an
essential ingredient for present and future global PDF analyses.
A wide range of data from the LHC for this process are available,
both for inclusive cross-sections~\cite{Collaboration:2010ey,
Chatrchyan:2013faa,Aad:2012mza}
and for differential distributions~\cite{Chatrchyan:2012saa,Aad:2012hg}, 
and much more will be measured in the near future.

As usual, the production of stable $t\bt$ pairs in \aNLO\ is simulated by
starting with the generation of the process. Since in the default model 
used by \aNLO\ there are four light flavours, one needs first to import a 
proper massless-bottom model. This is done
by means of the following commands:

\noindent
~~\prompt\ {\tt ~import model loop\_sm-no\_b\_mass}

\noindent
~~\prompt\ {\tt ~define p = g u d s c b u\~{} d\~{} s\~{} c\~{} b\~{}}

\noindent
Here, the second command instructs \aNLO\ that the proton
contains five light flavours, and does include the bottom quark;
by doing so, partonic subprocesses will be generated that feature
bottom quarks in the initial state. After this, one may proceed
with the generation command;

\noindent
~~\prompt\ {\tt ~generate p p > t t\~{} [QCD]} 

\noindent
followed by the standard {\tt output} and {\tt launch} commands
(see sect.~3 of ref.~\cite{Alwall:2014hca} for more details).
There are $n_l=7$ contributing parton luminosities. For each
of them, we report in table~\ref{ttlist}
\begin{table}[h]
{\small
\begin{center}
\begin{tabular}{ccccccc}
$l$  & $n_{rs}$  & $(r,s)$ & & & & \\
1 & 1 & $(g\,, g)$ & & & &  \\                                                                    
2 & 5 & $(\bb \,, g)$  & $(\cb \,, g)$  &   $(\bs \,, g)$ &    $(\ub \,, g)$  &   $(\db \,, g)$ \\
3 & 5 & $(d\,, g)$   & $(u\,, g)$  &  $(s\,, g)$ & $(c \,,g)$ & $(b \,,g)$ \\                     
4 & 5 & $(g\,, \bb)$  & $(g\,, \cb)$  &  $(g \,,\bs)$ &  $( g\,, \ub)$ &   $(g \,,\db)$ \\        
5 & 5 & $(g \,,d)$ & $(g \,,u)$  &  $(g \,,s)$  & $(g \,,c)$ & $(g \,,b)$ \\                      
6 & 5 & $(d \,,\db)$  &  $(u \,,\ub)$ &  $(s \,,\bs)$  & $(c \,,\cb)$ & $(b \,,\bb)$ \\           
7 & 5 & $(\bb \,,b)$ & $(\cb \,,c)$ &  $(\bs \,,s )$ &  $(\ub \,,u)$  & $(\db\,,d)$ \\            
\end{tabular}
\end{center}
}
\caption{\label{ttlist}
Values of $l$ and $(r,s)$ for which $T_{rs}^{(l)}=1$, as generated by
\amcfast\ in $t\bt$ production.
For a given $l$, the number $n_{rs}$ of non-null terms in the sum on
the r.h.s.~of eq.~(\ref{lumldef}) is also given.
}
\end{table}
\begin{figure}
 \begin{center}
 \epsfig{file=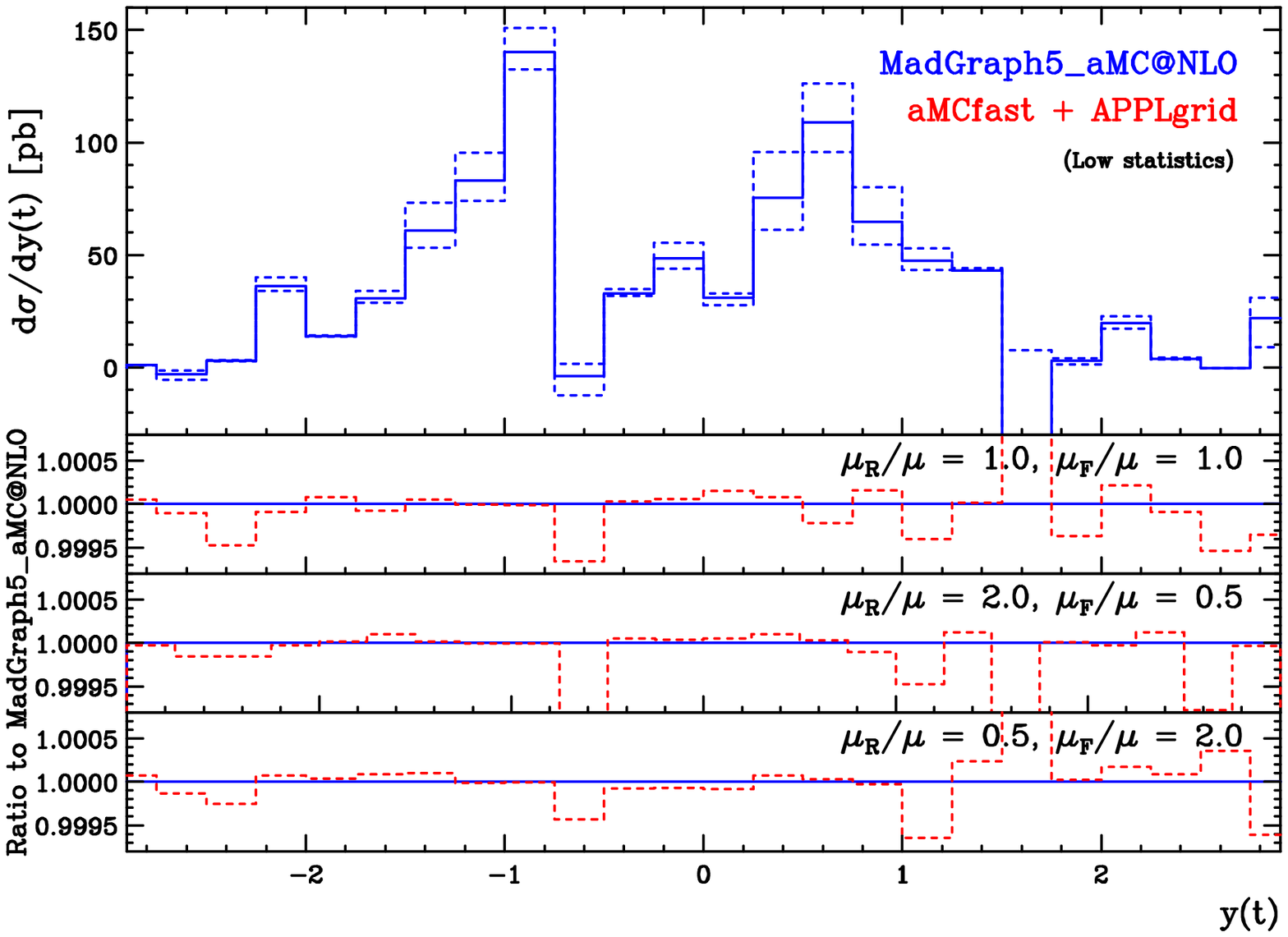, width=0.48\textwidth}
 \epsfig{file=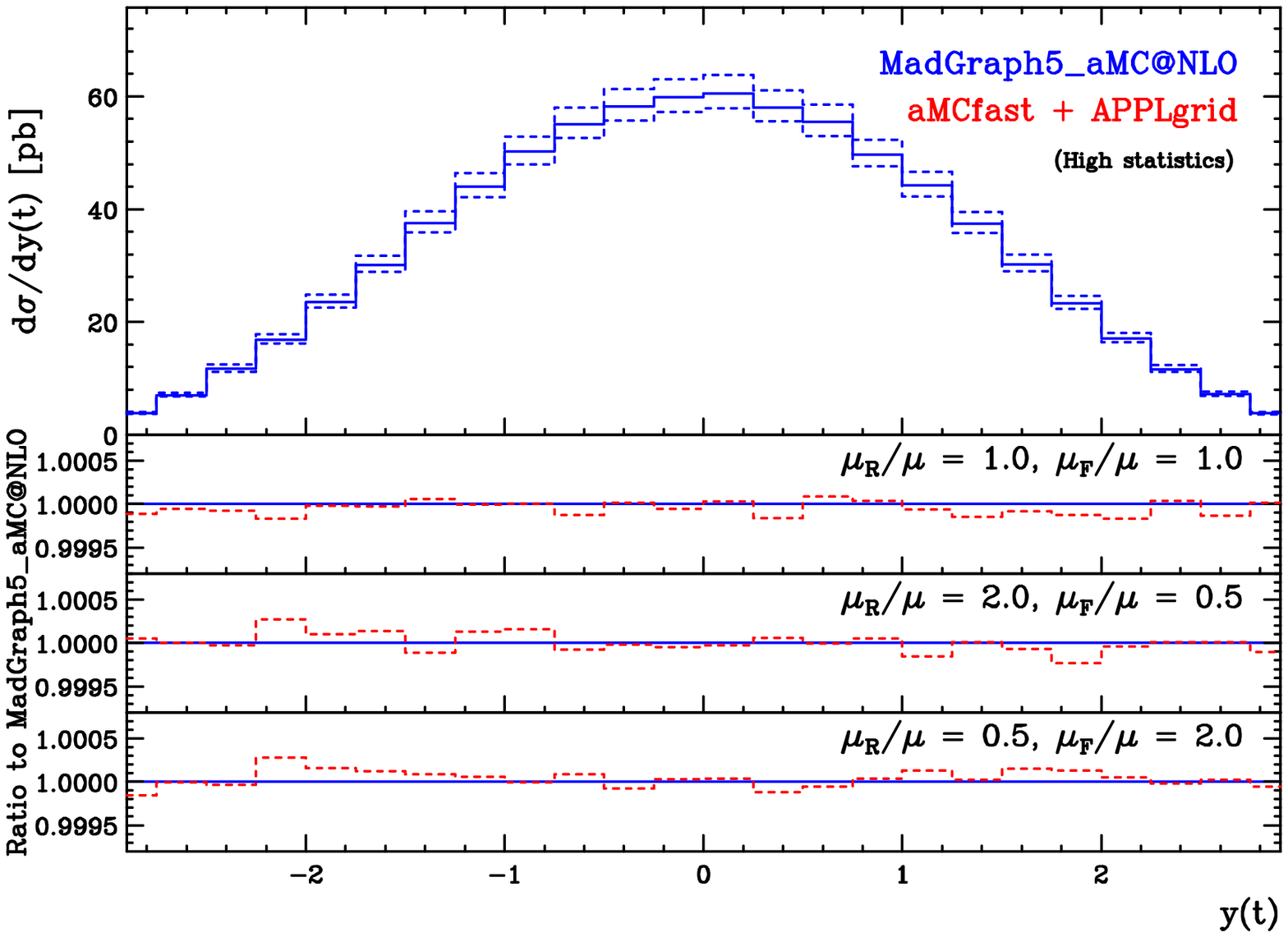, width=0.48\textwidth}
 \end{center}
\caption{Rapidity of the top quark, in $t\bt$ production
at the 14~TeV LHC, for a low- (left panel) and high-statistics
(right panel) run. In the main frames we show the reference \aNLO\ 
results obtained with central (eq.~(\ref{sc1}), solid histogram) and with 
non-central (eqs.~(\ref{sc2}) and~(\ref{sc3}), dashed histograms)
scale choices. The three lower insets present the ratios of the 
reconstructed results over the corresponding reference result,
for each of the three scale choices.
}
\label{fig:ytt_ttbar}
\end{figure}
the corresponding $(r,s)$ contributions, which are the terms on the 
r.h.s.~of eq.~(\ref{lumldef}) with $T_{rs}^{(l)}=1$; the number of
such terms, denoted by $n_{rs}$, is also reported in this table.
As was already mentioned in sect.~\ref{sec:int}, the assignments
of table~\ref{ttlist} are automatically determined by \amcfast\
using the information provided by \aNLO. 
We point out that each of the lines in table~\ref{ttlist} corresponds
to the set of four grids of eqs.~(\ref{appNLOz})--(\ref{appNLOB}).
It is possible to use them separately, e.g.~in order to determine 
the relative contribution of each parton luminosity to the different 
bins of the kinematical distributions studied,
although this feature has not been used in this paper.

\begin{figure}
 \begin{center}
 \epsfig{file=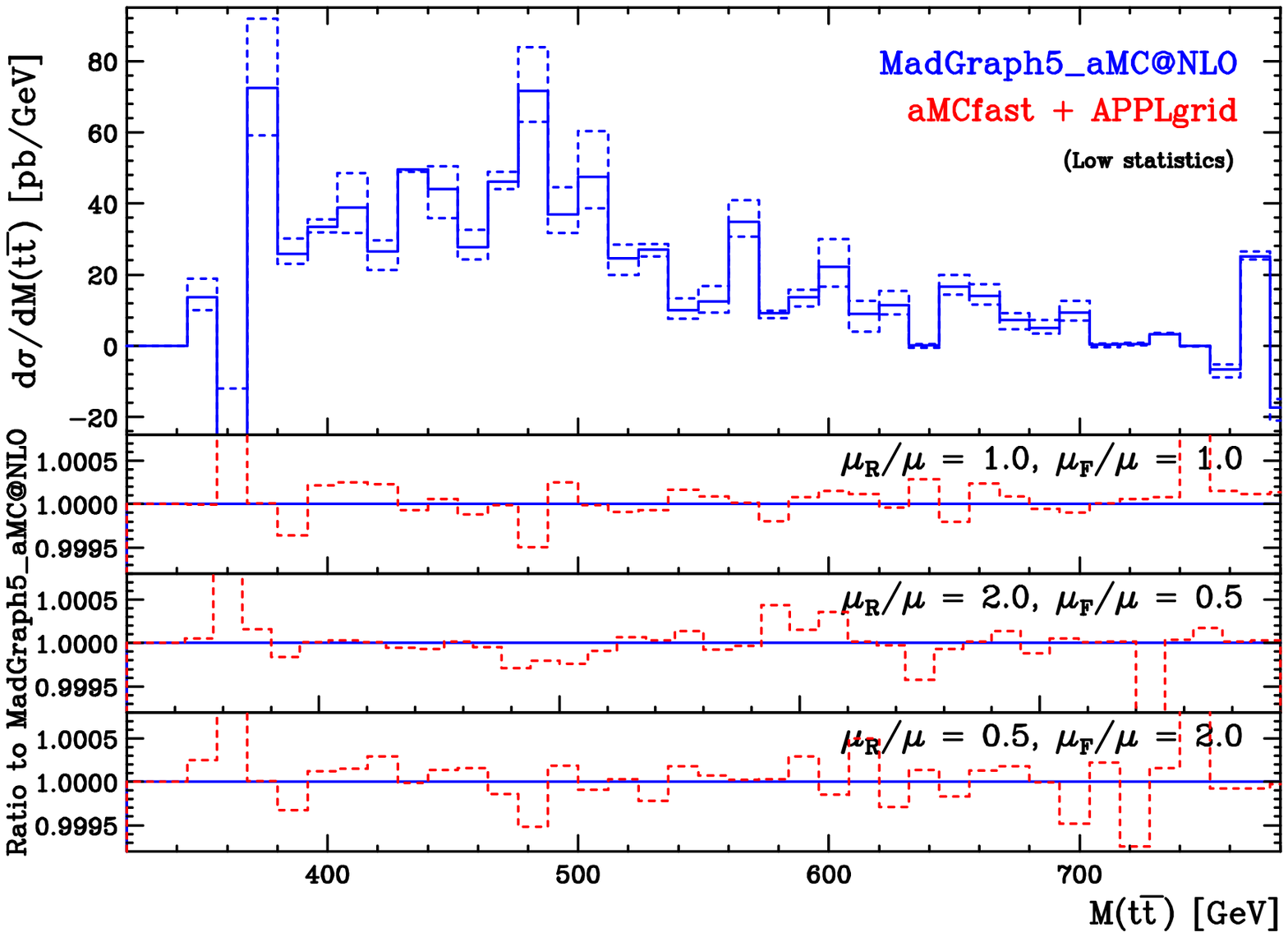, width=0.48\textwidth}
 \epsfig{file=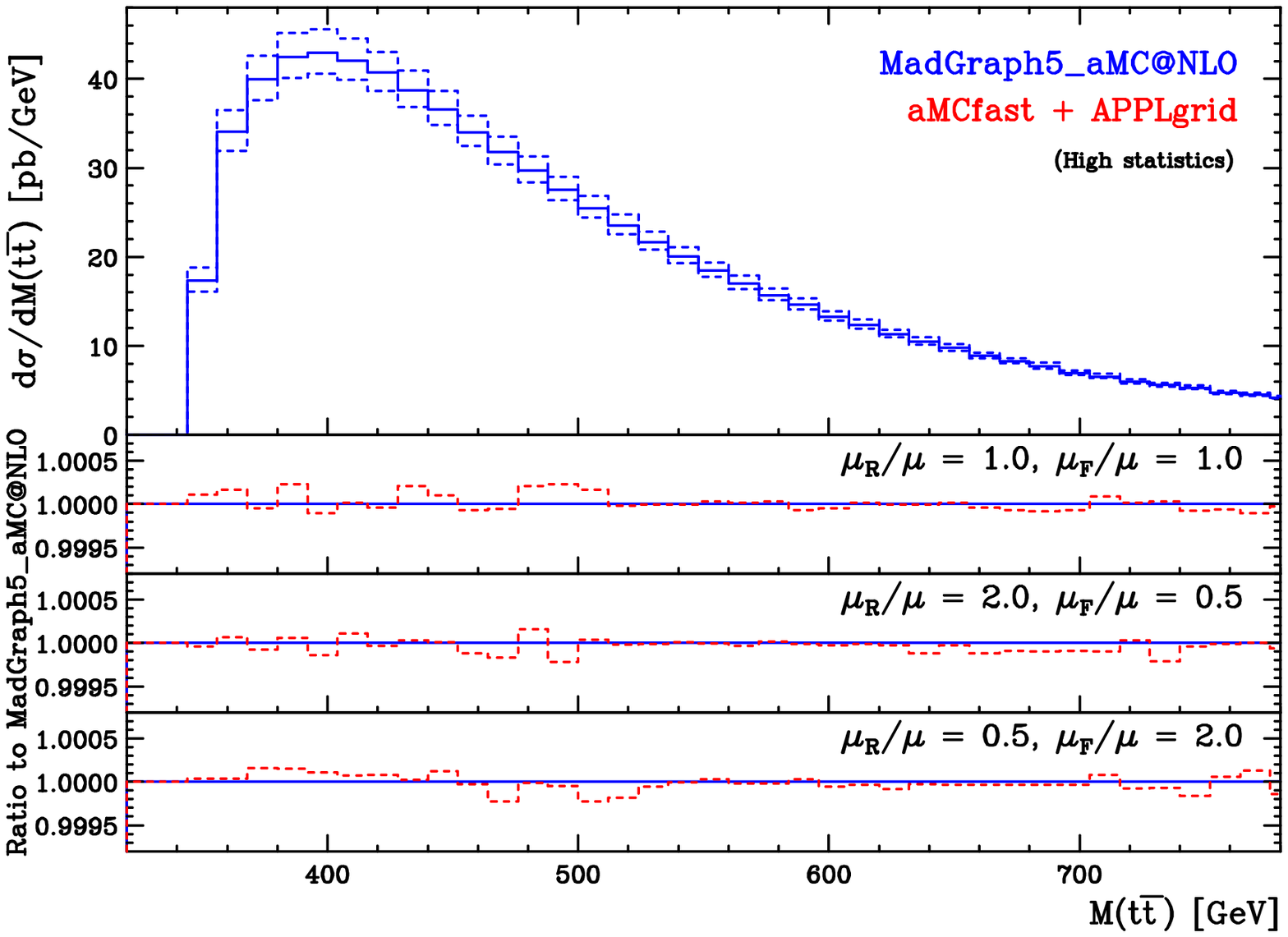, width=0.48\textwidth}
 \end{center}
\caption{As in fig.~\ref{fig:ytt_ttbar}, for the $t\bt$ invariant mass.
}
\label{fig:mtt_ttbar}
\end{figure}
The results of our validation are presented in figs.~\ref{fig:ytt_ttbar} 
and~\ref{fig:mtt_ttbar}, where we show the rapidity of the 
top quark ($y(t)$), and the $t\bt$ invariant
mass ($M(t\bt)$), respectively. Each figure consists
of two panels, with that of the left (right) obtained with a
low-statistics (high-statistics) run. Each panel has a main frame,
where the solid histogram is the \aNLO\ reference result obtained
with the central scales of eq.~(\ref{sc1}), and the dashed histograms
are the \aNLO\ reference results relevant to the two non-central
scale choices of eqs.~(\ref{sc2}) and~(\ref{sc3}) -- incidentally,
we note that the latter two predictions are obtained by using the
reweighting technique presented in ref.~\cite{Frederix:2011ss}
(i.e., \aNLO\ has been run only once). The three lower insets
show the accuracy obtained with interpolating grids, since they 
display as dashed histograms the ratio of the reconstructed over 
the reference result, for each of the three scale choices. These
histograms thus represent the validation of the \aNLO\ -- \amcfast\ 
-- \applgr\ chain, with values equal to one equivalent to a fast
computation of the cross section with zero interpolation errors.
As one can see from the plots, we obtain in all bins and for all scale
choices an accuracy of $3\cdot 10^{-4}$ at worst in the case of the
high-statistics run. In keeping with the general discussion given
at the beginning of sect.~\ref{sec:pheno}, the reconstruction accuracy
of the low-statistics results is also quite good, in spite of the 
fact that the corresponding differential cross sections display
extremely large fluctuations, and are unsuitable for phenomenology.

\subsection{Photon production in association with one jet\label{sec:gamma}}
The production of direct photons in hadronic collisions is 
sensitive to the gluon PDF, owing to the dominance of QCD 
Compton-scattering-like diagrams.
Thanks to tight isolation requirements applied to both theory
prediction and data, one is able to significantly reduce or even 
eliminate completely the cross section of non-direct photons, 
due to poorly-known fragmentation-fuction contributions 
in the former, and to the contaminating $\pi^0\to\gamma\gamma$
decays in the latter.
A recent re-analysis of all available isolated-photon collider data 
(inclusive in additional QCD radiation) has shown
that NLO QCD gives a good description of all the data sets,
and in addition it can provide one with some constraints on the gluon 
PDF in the region of Bjorken-$x$ relevant to the 
calculation of the gluon-fusion Higgs production cross
section~\cite{d'Enterria:2012yj}.
Therefore, there is a solid case to include photon data in the next
generation of global PDF analyses.
Data for isolated-photon production has been released by ATLAS and 
CMS~\cite{Aad:2013zba,Chatrchyan:2011ue,Khachatryan:2010fm};
in particular, the latest ATLAS measurement~\cite{Aad:2013zba}, 
that makes use of the complete 7 TeV run statistics, extends their 
kinematical coverage in the photon transverse energy up to a value
of $\Et(\gamma)\sim 600$ GeV (see also ref.~\cite{ATL-PHYS-PUB-2013-018} 
for a study of the sensitivity of this measurement to various sets of 
PDFs).

On top of inclusive isolated-photon production, that of isolated photons 
in association with extra jets allows one to extend the sensitivity to 
the gluon PDF to a wider range of Bjorken-$x$, and also provides one
with some information on the quark PDFs. Some measurements of isolated
photon plus jets have been reported by ATLAS and CMS~\cite{Aad:2013gaa,
Chatrchyan:2013oda,Chatrchyan:2013mwa}; while these data are still not 
precise enough to directly constrain PDFs~\cite{Carminati:2012mm}, 
the increase of the statistics collected in future LHC runs will
dramatically change the picture, so that this process will become
a useful addition to global PDF fits.

We have thus computed $\gamma+$jet production with \aNLO.
The computation is performed with five light flavours, so the generation
command must be preceded by the same {\tt import} and {\tt define}
commands which have been used in the case of $t\bt$ production (see
sect.~\ref{sec:tt}). On top of those, we must also use here:

\noindent
~~\prompt\ {\tt ~define j = g u d s c b u\~{} d\~{} s\~{} c\~{} b\~{}}

\noindent
since a light jet must have the same parton contents as the proton.
The generation command is then:

\noindent
~~\prompt\ {\tt ~generate p p > a j [QCD]} 

\noindent
We have imposed the photon to be hard and central,
$\pt(\gamma)\ge 80$~GeV and $\abs{\eta(\gamma)}\le 3$.
We use the Frixione photon-isolation prescription~\cite{Frixione:1998jh},
since in this way one sets identically equal to zero the contribution of the
fragmentation-function component in an infrared-safe manner.
\begin{figure}
 \begin{center}
 \epsfig{file=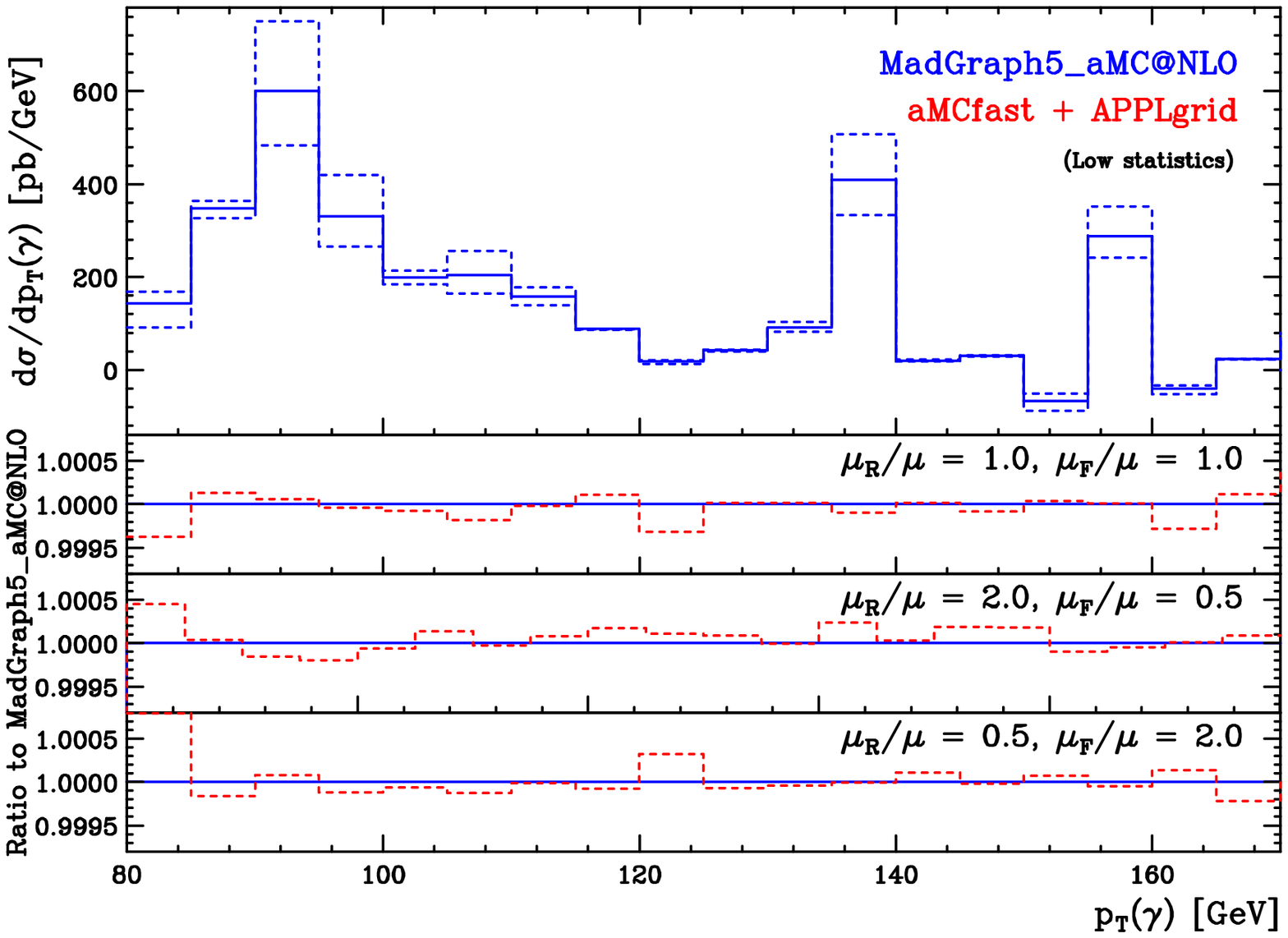, width=0.48\textwidth}
 \epsfig{file=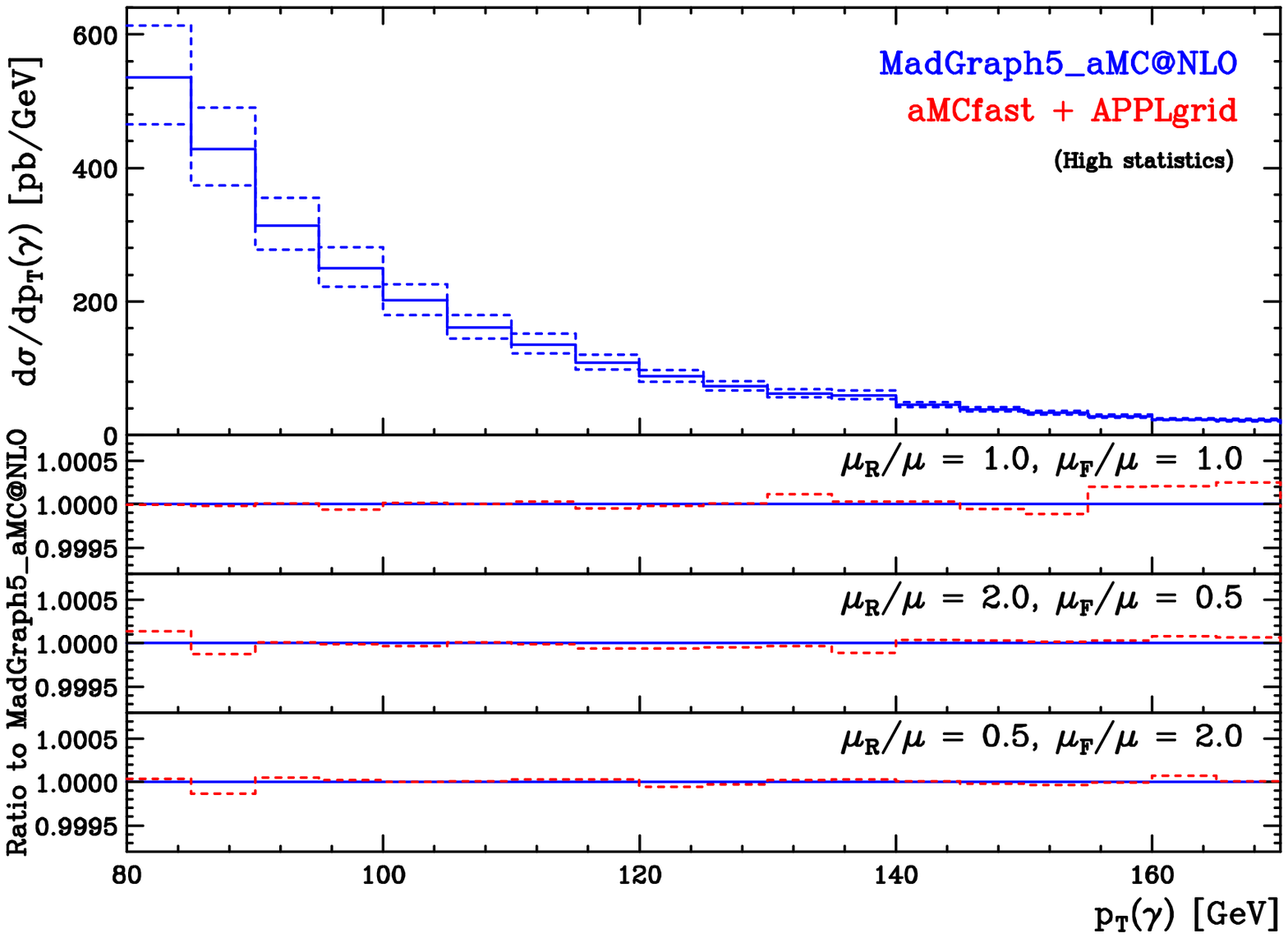, width=0.48\textwidth}
 \end{center}
\caption{As in fig.~\ref{fig:ytt_ttbar}, for the photon transverse momentum,
in the hadroproduction of an isolated photon in association with an extra jet.
}
\label{fig:ptgamma_gammajet}
\end{figure}
The parameters of the isolation (which in \aNLO\ can be easily
modified at the run-card level) are set as follows:
$\epsilon_{\gamma}=1.0$ and $n=1$ (see eq.~(3.4) of 
ref.~\cite{Frixione:1998jh}; note that, in an hadronic environment,
the energies and angles of that equation have to be formally replaced
by transverse energies and distances in the $(\eta,\varphi)$ plane);
we use an isolation cone of radius $R_0=0.4$.
Jets are reconstructed with the anti-$\kt$ algorithm~\cite{Cacciari:2008gp},
as implemented in {\sc\small FastJet}~\cite{Cacciari:2011ma}, with a jet radius
of $R=0.4$, and subject to the conditions $\pt(j)\ge 30$~GeV and
$\abs{\eta(j)}\le 4.4$.
\begin{figure}
 \begin{center}
 \epsfig{file=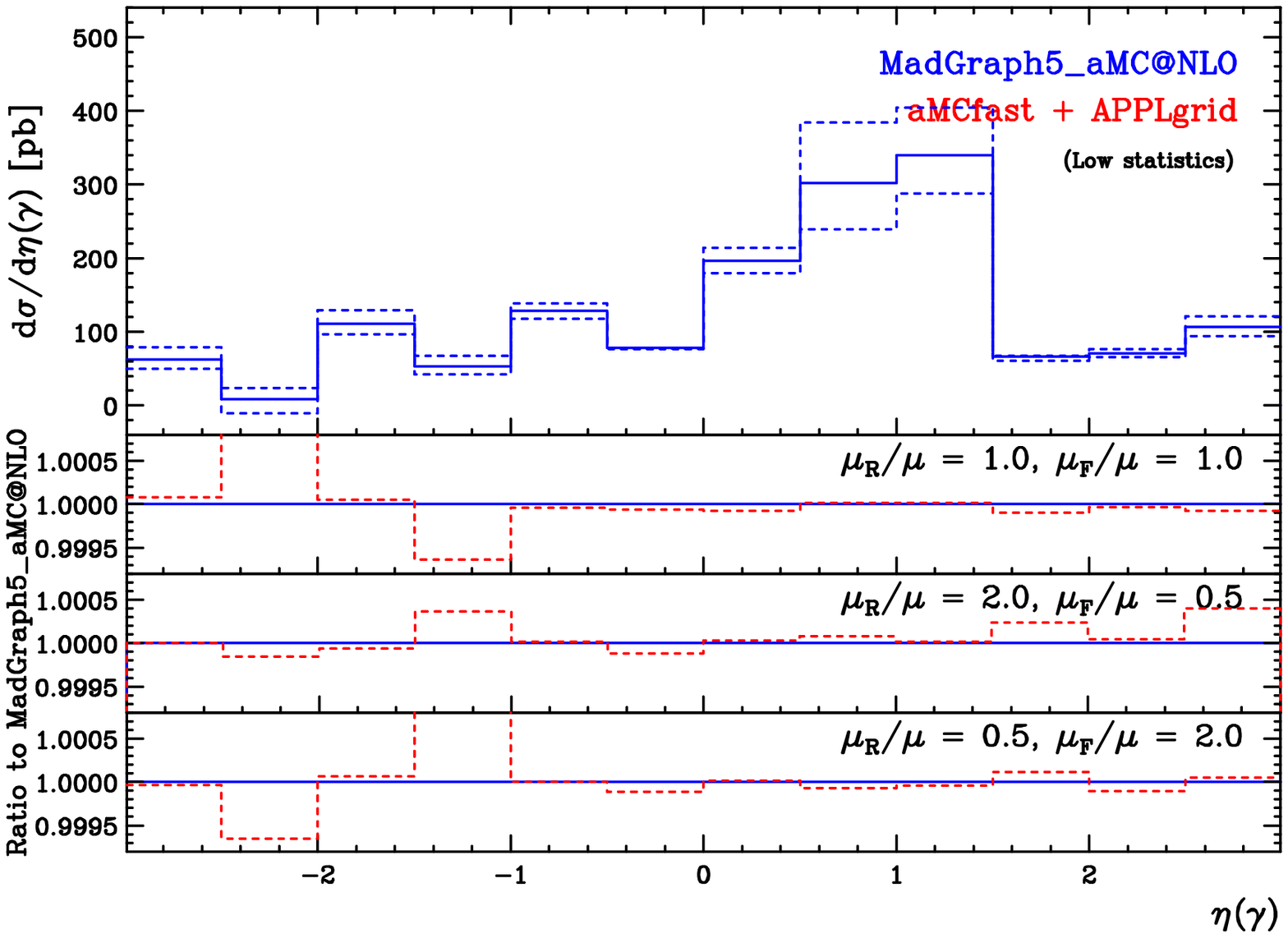, width=0.48\textwidth}
 \epsfig{file=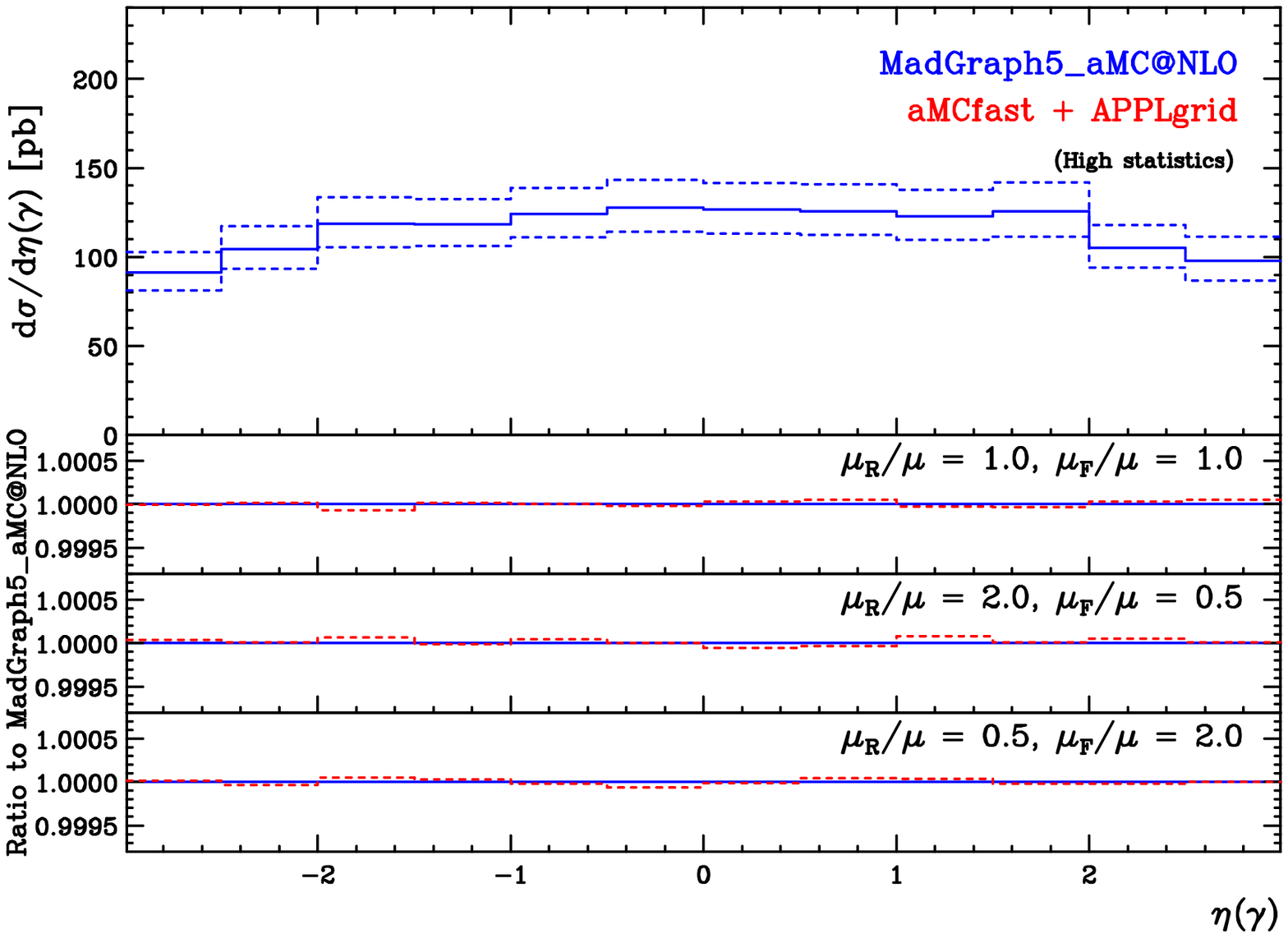, width=0.48\textwidth}
 \end{center}
\caption{As in fig.~\ref{fig:ptgamma_gammajet}, for the photon pseudorapidity.
}
\label{fig:etadist_gammajet}
\end{figure}
There are $n_l=33$ parton luminosities that contribute to the
current process; their list is given in table~\ref{pjlist}
(see appendix~\ref{sec:lum}).
That table shows that the definition of the parton luminosities in
\aNLO\ (i.e., eq.~(\ref{fact5})) is not maximally efficient. For
example, the matrix elements associated with $l=30$ and $l=31$ in
table~\ref{pjlist} are not equal, but differ by a trivial overall
factor (the charge squared of a up-type vs an down-type quark).
The combination of these two contributions could therefore be
achieved by simply relaxing the condition that all non-zero
$T_{rs}^{(l)}$ elements be equal to one. While this can easily
be done in the context of a specific computation, its general
implementation in an automated code able to deal with any user-defined
process is not straightforward.

The validation plots, presented in figs.~\ref{fig:ptgamma_gammajet} 
and~\ref{fig:etadist_gammajet}, follow the same pattern as those
for $t\bt$ production. The representative observables chosen here 
are the photon transverse momentum ($\pt(\gamma)$, 
fig.~\ref{fig:ptgamma_gammajet}) and the photon
pseudorapidity ($\eta(\gamma)$, fig.~\ref{fig:etadist_gammajet}).
As in sect.~\ref{sec:tt}, the agreement between the
reference and reconstructed results is at the level of 
$10^{-4}$ or better for the high-statistics run, and only slightly
worse for the low-statistics one.

\subsection{Dilepton production in association with one jet\label{sec:Zj}}
We now turn to studying the hadroproduction of a lepton pair in association
with one jet; the leptons originate from an intermediate virtual $Z$
or $\gamma$ boson, and we shall denote the pair with the shorthand notation 
$Z^\star$ in what follows. Inclusive $Z$-boson production has been used for 
a long time in PDF fits, since it provides one with a clean handle on the 
quark flavour separation in the proton, and various quality measurements from 
the Tevatron~\cite{Aaltonen:2010zza,Abazov:2007jy}
and the LHC~\cite{Aad:2011dm,Chatrchyan:2011wt} are available.
On the other hand, by requiring the presence of an additional extra jet
one enters in a different domain, since at the Born level one is
dominated by $qg$ channels, which thus allow one to probe directly
the gluon PDF. In this context, it is convenient to study the
large-$\pt(Z^\star)$ region: in fact, the fraction of gluon-initiated events 
increases\footnote{This comment applies to an intermediate hardness
region; asymptotically, $qg$-initiated contributions are strongly
suppressed by the fastly decreasing gluon PDF at large $x$'s.}
with $\pt(Z^\star)$, and furthermore by doing so one is safely
away from the small-$\pt(Z^\star)$ region, where fixed-order calculations
do not give physically-meaningful predictions.
Another advantage associated with this kinematic region is that ratios 
of cross sections for $W$ over $Z$ production in association with jets
can result in clean constraints on quarks and antiquarks PDFs at 
large-$x$~\cite{Malik:2013kba}, where they are poorly known (and especially
so for the antiquarks).

There are various ATLAS and CMS measurements available for $Z$+jets 
production, which typically focus on the comparison between theory and
data as a test of perturbative-QCD predictions which are capable of
describing a significant amount of hard radiation~\cite{Aad:2013ysa,
Aad:2011qv,Chatrchyan:2013tna,CMS-PAS-SMP-12-017}.
However, from the PDF point of view, it is more interesting to 
perform the analysis as inclusively as possible (bar for the first 
jet, which one needs in order to be sensitive to the gluon PDF,
as mentioned above), and to concentrate on the precise
measurement of the $Z^\star$ transverse momentum, possibly in bins 
of different rapidity $y(Z^\star)$. It is particularly convenient
to perform such a measurement with a leptonically-decaying $Z$,
which offers a cleaner scenario w.r.t.~that of a hadronic decay;
there are ongoing analyses in ATLAS and CMS --
see e.g.~\cite{CMS-PAS-SMP-13-013}.

We have computed dilepton$+$jet production at fNLO with \aNLO\ 
in the five-flavour scheme, following the generation 
command\footnote{The muons are treated as massless particles.}:

\noindent
~~\prompt\ {\tt ~generate p p > mu+ mu- j [QCD]} 

\noindent
which has been preceded by the same {\tt import} and {\tt define}
commands relevant to sect.~\ref{sec:gamma}. We have also imposed
the following cuts:
\beq
\pt(\mu^\pm)\ge 10~{\rm GeV}\,,\;\;\;\;\;\;
\abs{\eta(\mu^\pm)}\le 2.5\,,\;\;\;\;\;\;
M(\mu^+\mu^-)\ge 20~{\rm GeV}\,.
\eeq

The list of parton luminosities relevant to this process 
is the same as that for $\gamma+$jet production, and is given
in table~\ref{pjlist}. As in the $\gamma+$jet case,
jets are reconstructed by using the anti-$k_t$ algorithm with 
$R = 0.4$, $\pt(j)\ge 30$~GeV, and $\abs{\eta(j)}\le 4.4$.

\begin{figure}
 \begin{center}
 \epsfig{file=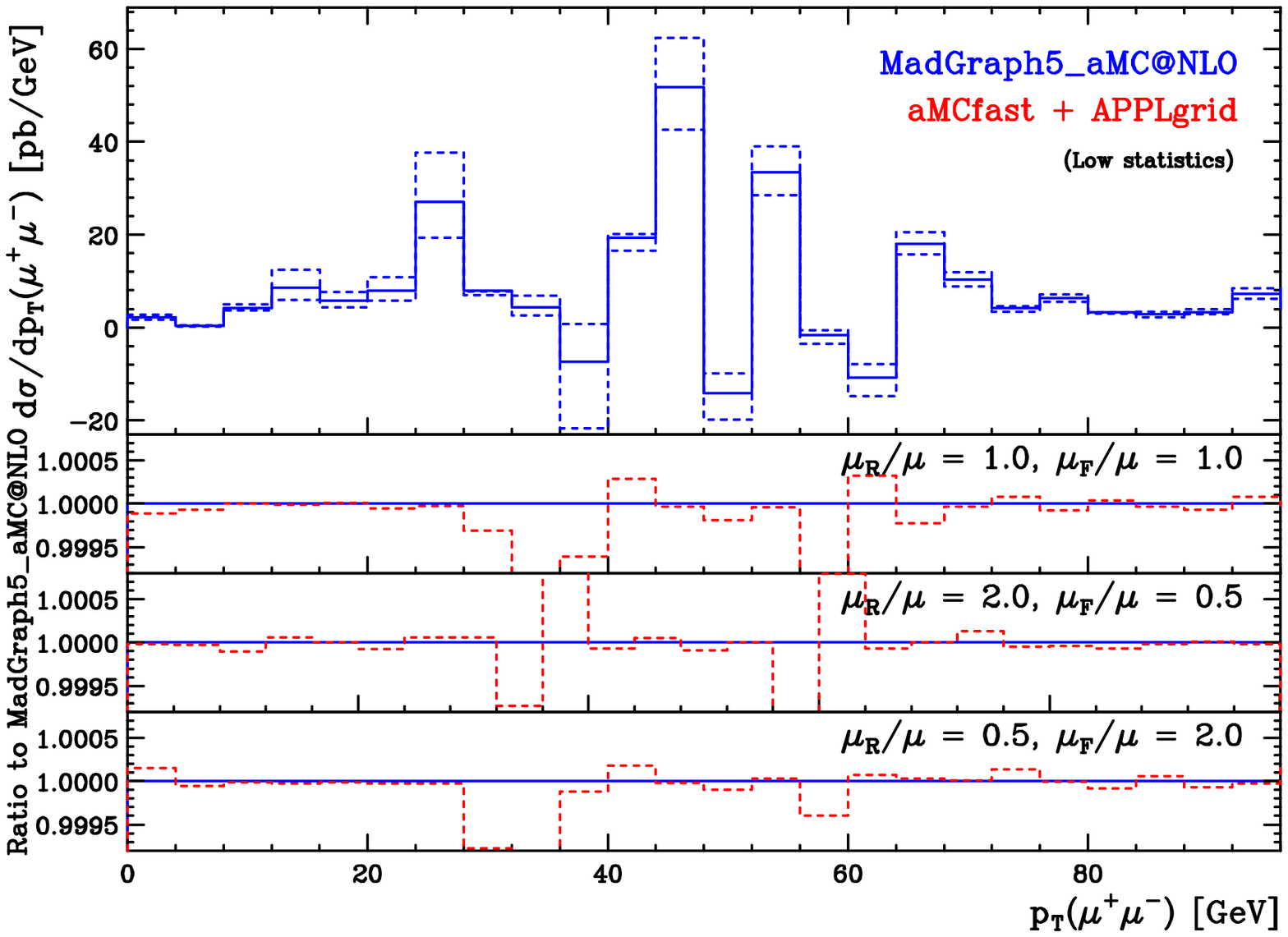, width=0.48\textwidth}
 \epsfig{file=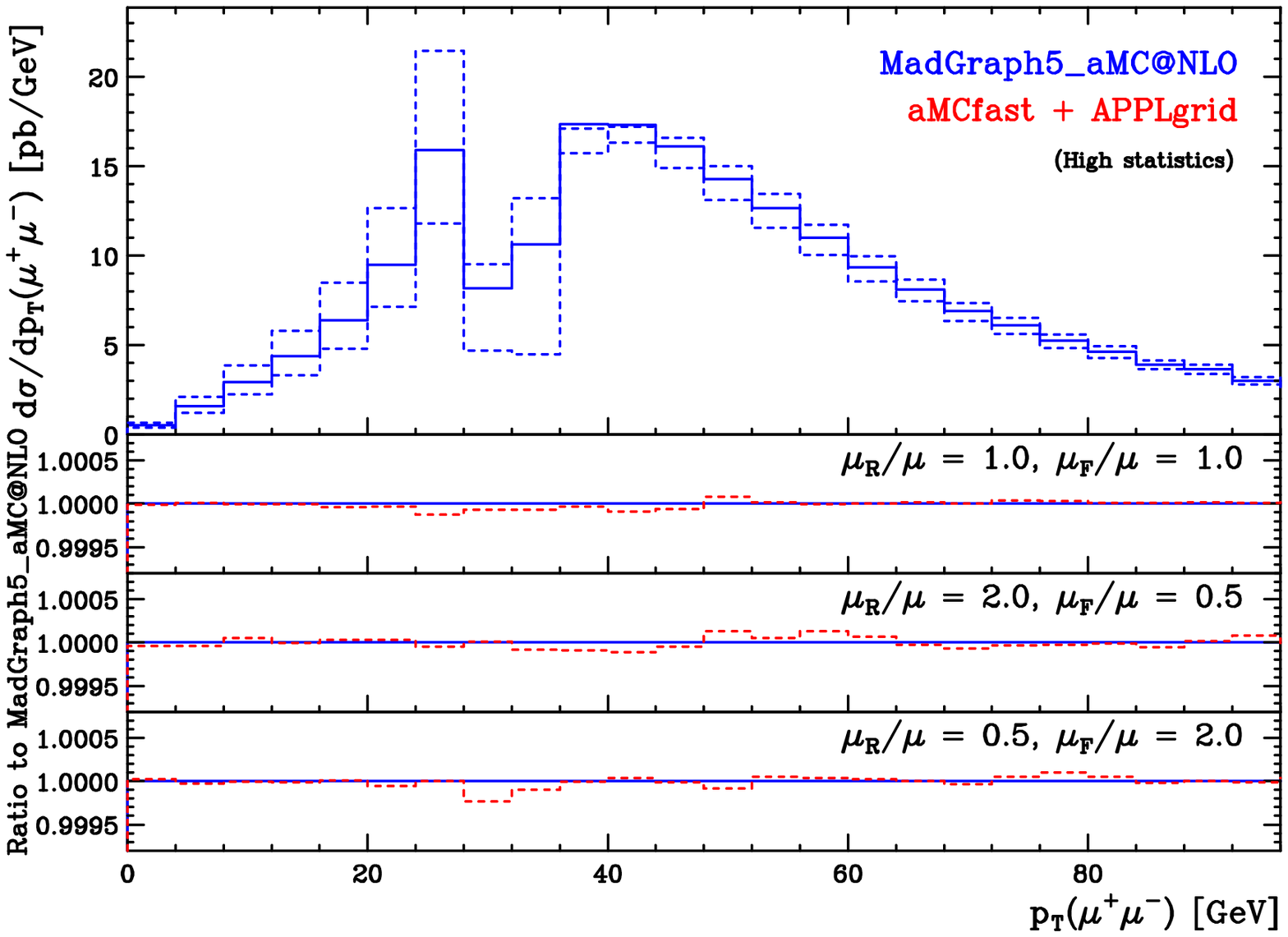, width=0.48\textwidth}
 \end{center}
\caption{As in fig.~\ref{fig:ytt_ttbar}, for the lepton-pair transverse 
momentum, in the hadroproduction of a lepton pair in association with an 
extra jet.
}
\label{fig:ptz_zjet}
\end{figure}
\begin{figure}
 \begin{center}
 \epsfig{file=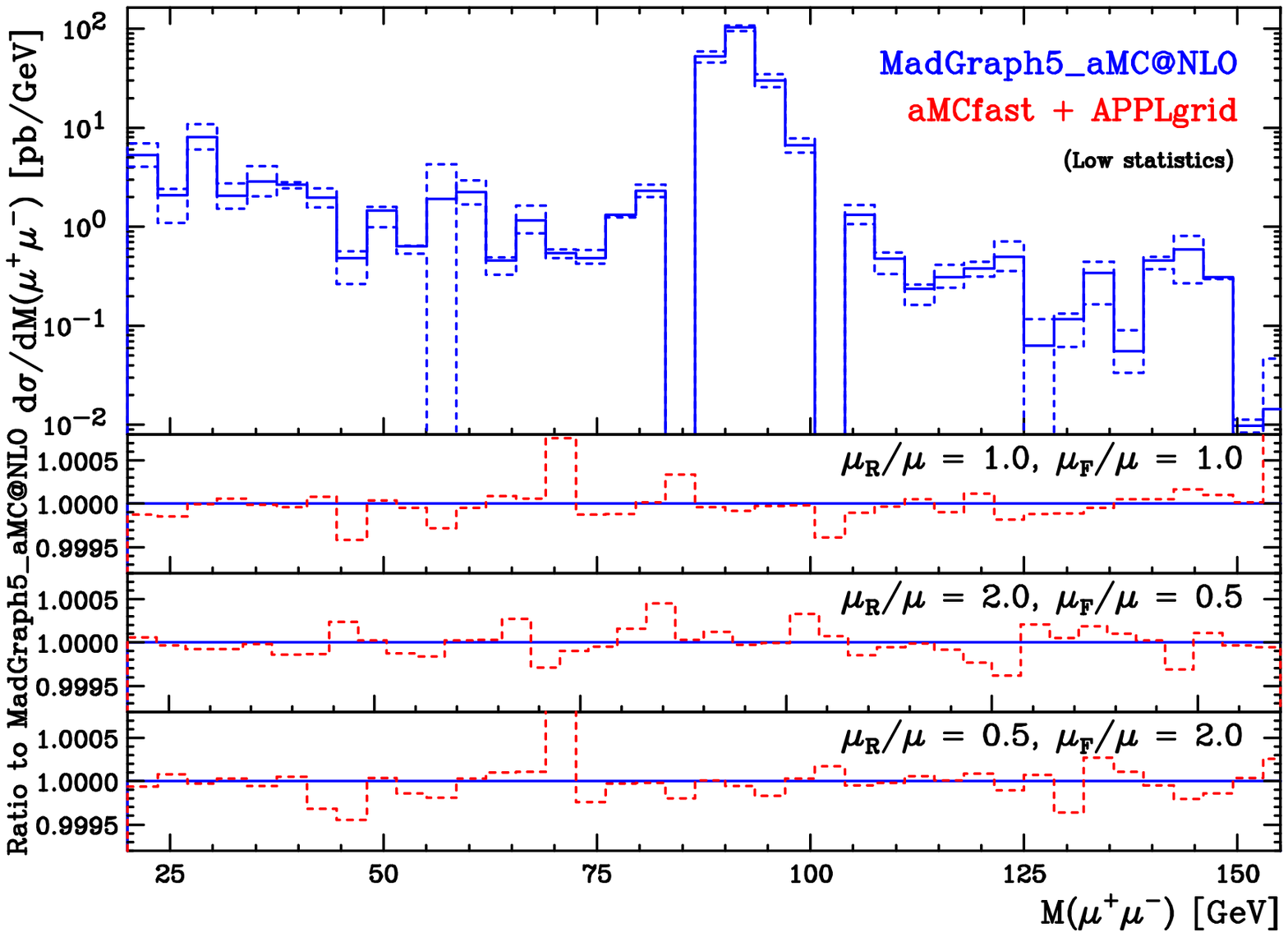, width=0.48\textwidth}
 \epsfig{file=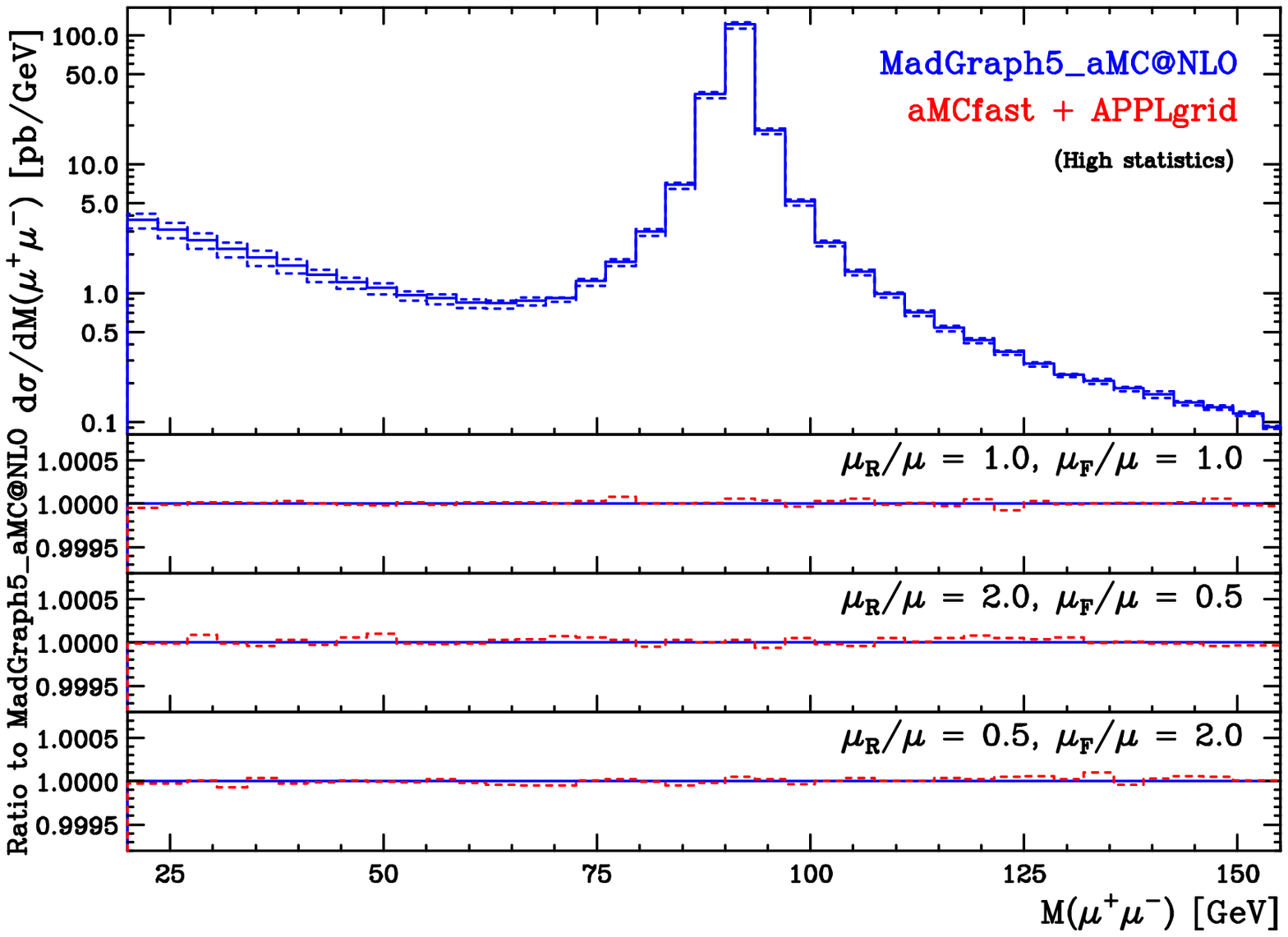, width=0.48\textwidth}
 \end{center}
\caption{As in fig.~\ref{fig:ptz_zjet}, for the lepton-pair invariant mass.
}
\label{fig:etaz_zjet}
\end{figure}
For the \amcfast\ validation we have chosen the transverse momentum
of the lepton pair and its invariant mass, $\pt(\mu^+\mu^-)$ and 
$M(\mu^+\mu^-)$, which we display in figs.~\ref{fig:ptz_zjet} 
and~\ref{fig:etaz_zjet} respectively. The latter clearly shows
the small-$M$ enhancement due to the presence of an intermediate photon.
In the former, the feature around $\pt(\mu^+\mu^-)\sim 30$~GeV is due
to the presence of a sharp kinematic threshold only at the Born level
(induced by the $\pt(j)$ cut), which causes this region to be 
infrared-sensitive~\cite{Catani:1997xc}. 
Although this is an issue from the physics point of
view for perturbative calculations, it does not pose any problems
for the reconstruction through interpolating grids.
We observe again an excellent agreement between the reference
and the reconstructed results.

\subsection{$Z$ production in association with a 
$b\bar{b}$ pair\label{sec:Zbb}}
The production of a $Z$ boson in association with a bottom-antibottom quark
pair is an interesting example of how the automation of a fast interface to 
NLO QCD computations achieved by \amcfast\ can be a valuable tool to easily
include arbitrarily complicated process into a global PDF fit.  A related
application could be that of studying to which degree the recent 
discrepancies between theory and experiment 
(see e.g.~ref.~\cite{Chatrchyan:2014dha}) for the production of $Z$ 
bosons in association with bottom quarks might be absorbed by a change
in PDFs. The hadroproduction of $Wb\bb$ and $(Z/\gamma^*)b\bb$ in the \aNLO\ 
framework has been previously discussed in ref.~\cite{Frederix:2011qg},
without however considering any aspects relevant to PDFs.

The generation of the current process in \aNLO\ is achieved through
the command:

\noindent
~~\prompt\ {\tt ~generate p p > z b b\~{} [QCD]} 

\noindent
Note that the default model adopted by \aNLO\ for NLO computations treats 
the bottom quarks as massive particles, and one works in a four-flavour 
scheme: hence, no models are explicitly imported, and no multi-particle 
labels redefined, at variance with was 
done previously. This is important in order to
be able to perform the computation of $Zb\bb$ production without
resorting to any kind of jet-reconstruction algorithm, to probe
the $b$ quarks down to zero transverse momentum, and to define a
fully-inclusive cross section.
The list of partonic luminosities that contribute to this process 
is given in table~\ref{zbblist}.
\begin{figure}
 \begin{center}
 \epsfig{file=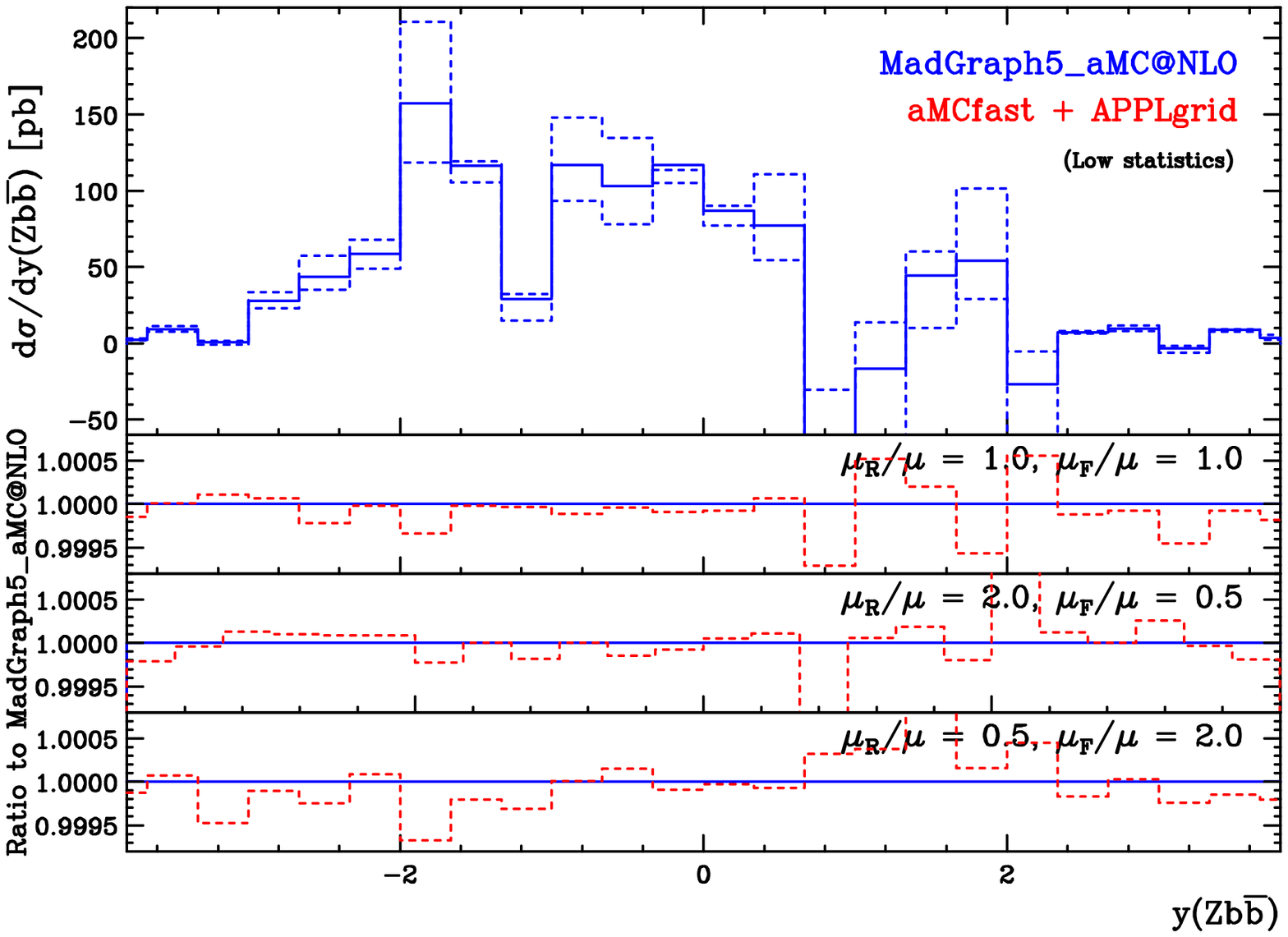, width=0.48\textwidth}
 \epsfig{file=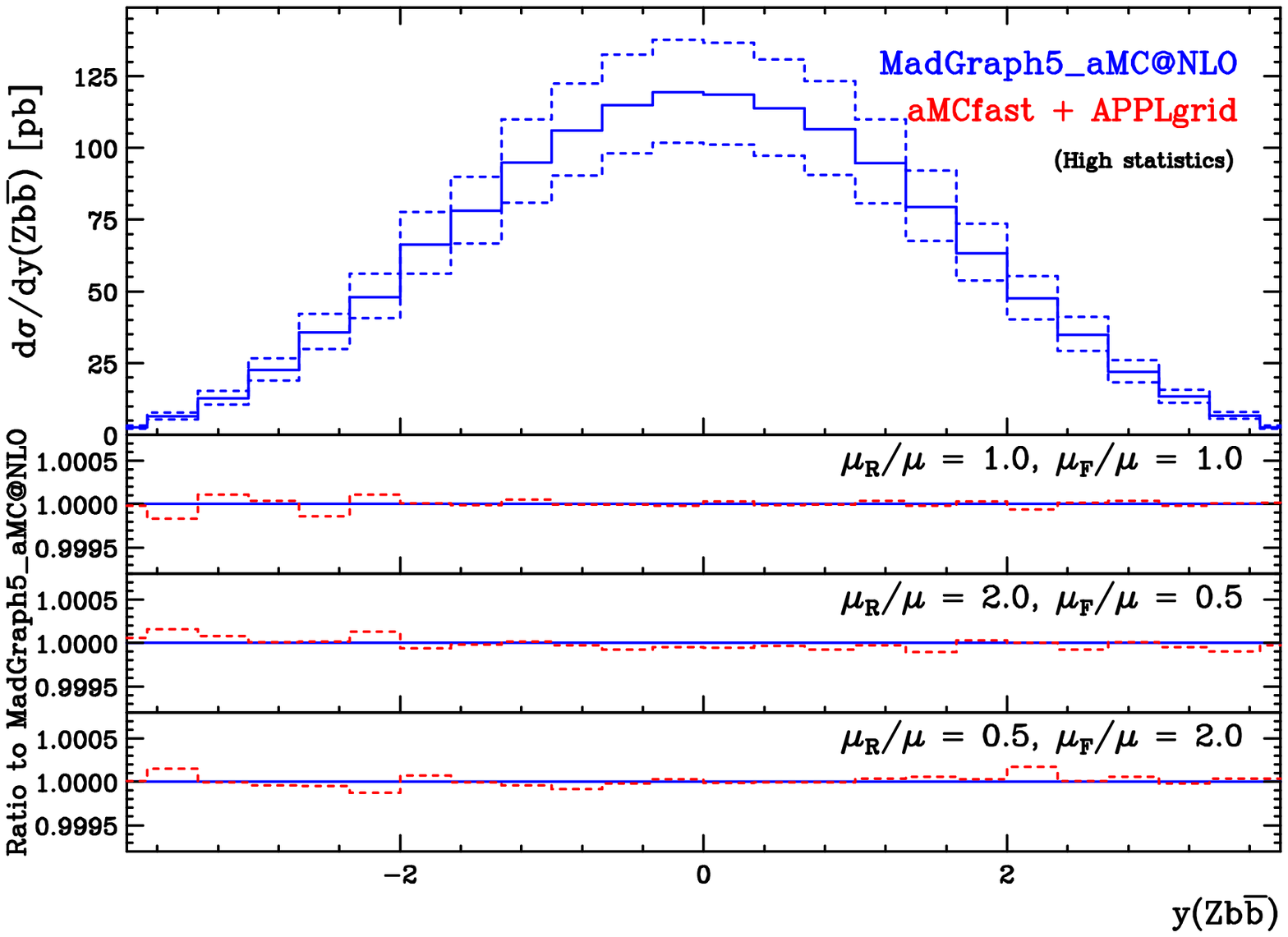, width=0.48\textwidth}
 \end{center}
\caption{As in fig.~\ref{fig:ytt_ttbar}, for the rapidity of the $Zb\bb$ 
system, in $Zb\bb$ hadroproduction.
}
\label{fig:yzbb_zbb}
\end{figure}
We validate the results of the present simulation by considering
the rapidity distribution of the $Zb\bb$ system (shown in 
fig.~\ref{fig:yzbb_zbb}), and the transverse momentum of the
$b\bb$ pair (shown in fig.~\ref{fig:mzbb_zbb}). As in the previous cases, 
we observe no significant deviations of the reconstructed from the reference
predictions.

\subsection{$W$ production in association with charm quarks\label{sec:Wc}}
The final representative example of an LHC process of interest for PDF fits
that we discuss in this work is the hadroproduction of $W$ bosons in 
association with charm quarks. This process is directly sensitive to the 
PDF of the strange quark (see e.g.~refs.~\cite{Baur:1993zd,Stirling:2012vh}), 
which among the light-quark PDFs is the most poorly determined.
In the majority of global PDF fits, the strange PDF is constrained 
by neutrino data~\cite{MasonPhD,Mason:2007zz,Goncharov:2001qe}
(in particular by the so-called dimuon process,
i.e.~charm production in charged-current DIS).
The use of neutrino DIS data in PDF fits has several drawbacks, such as the 
need to model charm fragmentation, and to understand charm-quark
mass effects at low scales.
\begin{figure}
 \begin{center}
 \epsfig{file=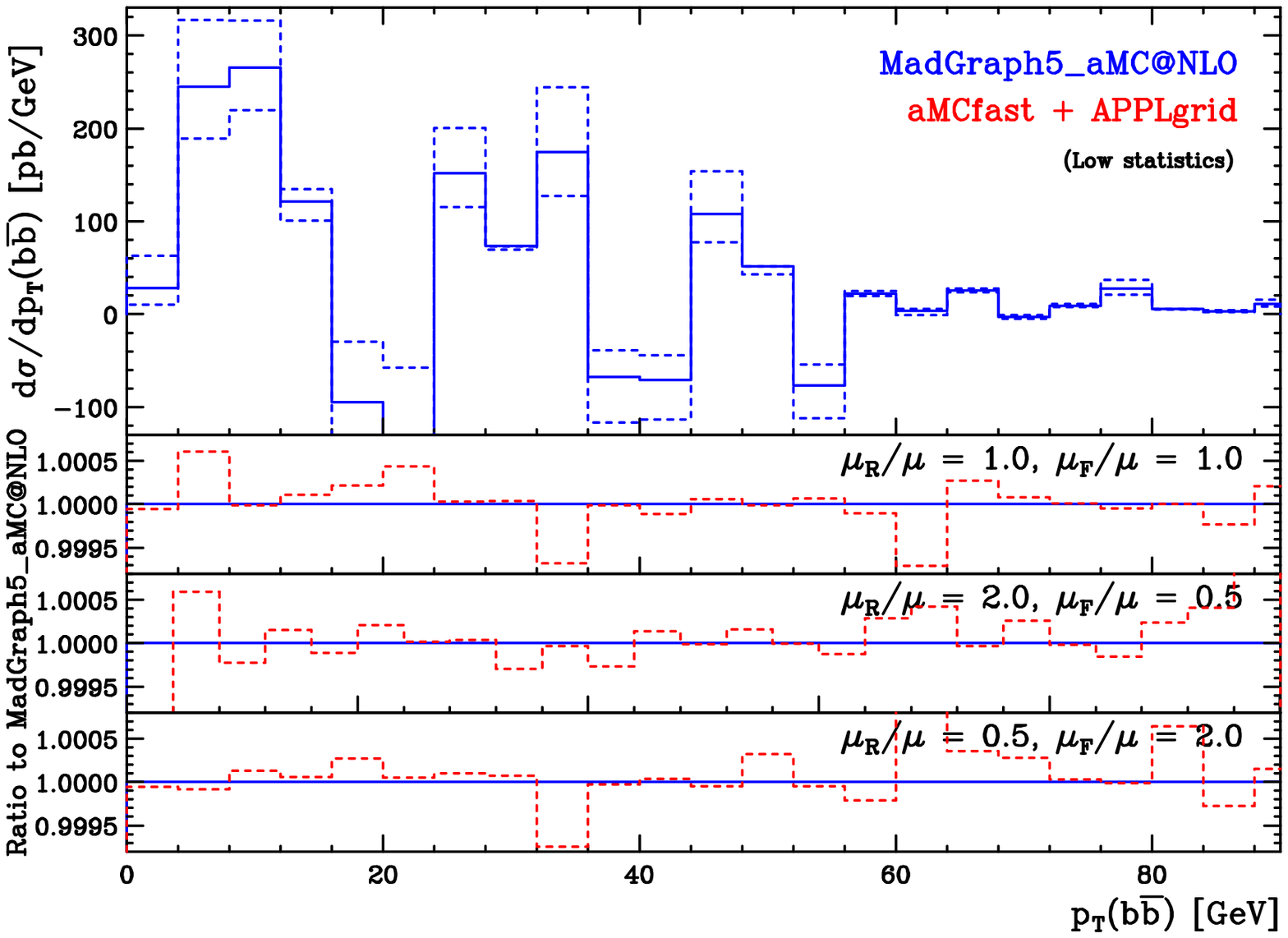, width=0.48\textwidth}
 \epsfig{file=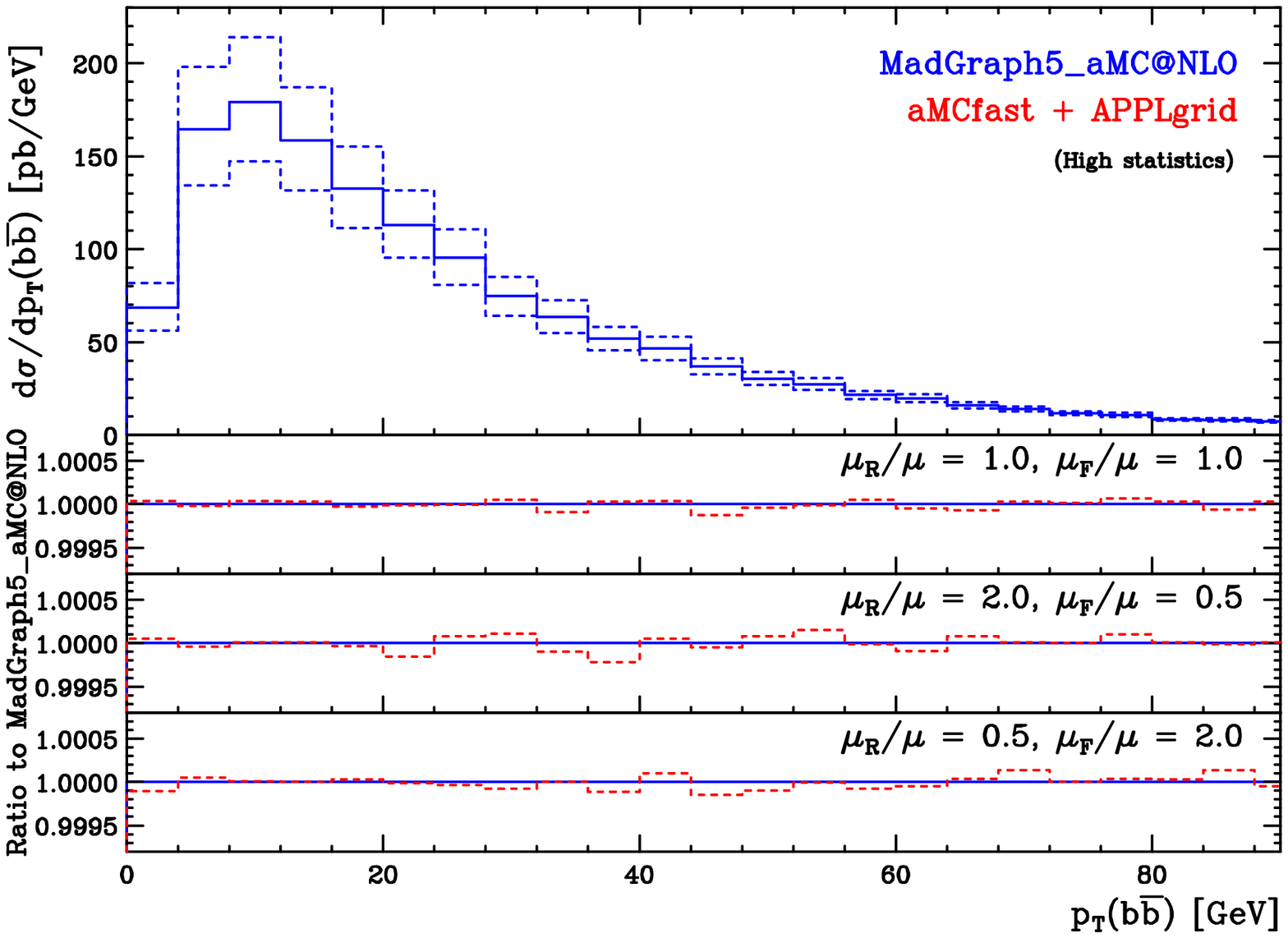, width=0.48\textwidth}
 \end{center}
\caption{As in fig.~\ref{fig:yzbb_zbb}, for the transverse momentum 
of the $b\bb$ pair.
}
\label{fig:mzbb_zbb}
\end{figure}

It is therefore quite interesting that $Wc$ hadron-collider data 
provide one with a clean and robust independent check
of the strange PDF determined elsewhere. In addition to that, 
the differences in the production rates of $W^+$ and $W^-$ bosons
in association with charm quarks give information on the strange asymmetry 
in the proton, that plays an important role in the 
explanation~\cite{Ball:2009mk} of the NuTeV anomaly~\cite{Goncharov:2001qe}.
Data on $W$ production in association with either charm jets or $D$ mesons
have been made public by ATLAS~\cite{Aad:2014xca} 
and CMS~\cite{Chatrchyan:2013uja}, based on the 7 TeV
2011 data sets, and analogous measurements from the 8 TeV run are ongoing.
In particular, the CMS $Wc$ data has been recently used in a QCD 
analysis~\cite{Chatrchyan:2013mza}, together with HERA data, to show that
the strange-quark PDF from collider-only data can be determined with 
a precision comparable to that of global fits which include neutrino data.
Another recent analysis of the compatibility of the LHC $Wc$ data with 
existing fixed-target DIS and Drell-Yan data has been presented in 
ref.~\cite{Alekhin:2014sya}.

Once again, we now compare a couple of reference distributions from 
\aNLO\ to those obtained with the a-posteriori \amcfast\ -- \applgr\ 
convolution. For simplicity, we restrict ourselves to considering 
$W^+\bar{c}$ production; on the other hand, the process we shall
actually deal with is that where an electron and a neutrino, rather
than a $W$, are present in the final state, so as to make sure
that all production spin correlations are correctly taken into account. 
Incidentally, this also shows that one is not restricted to performing
computations in the narrow width approximation; in particular,
the complex mass scheme~\cite{Denner:2005fg} is fully supported in \aNLO.
Similarly to the case of the $b$ quark in $Zb\bb$ production discussed 
in sect.~\ref{sec:Zbb}, in the present case it is best to treat the charm
as a massive particle. Since in the default model used by \aNLO\ the
charm is massless, one needs to import a proper massive-charm model
before proceeding with the generation of the process. This is done
by means of the following commands:

\noindent
~~\prompt\ {\tt ~import model loop\_sm-c\_mass}

\noindent
~~\prompt\ {\tt ~define p = g u d s u\~{} d\~{} s\~{}}

\noindent
~~\prompt\ {\tt ~generate p p > e+ ve c\~{} [QCD]} 

\noindent
In this list, the second command instructs \aNLO\ that the proton
contains three light flavours, and does not feature the charm quark;
by doing so, no partonic subprocesses will be generated that feature
a charm in the initial state. The following cuts have been imposed:
\beq
\pt(e^+)\ge 10~{\rm GeV}\,,\;\;\;\;\;\;
\abs{\eta(e^+)}\le 2.5\,.
\eeq
The list of partonic luminosities that contribute to this process 
in given in table~\ref{wclist}.

\begin{figure}
 \begin{center}
 \epsfig{file=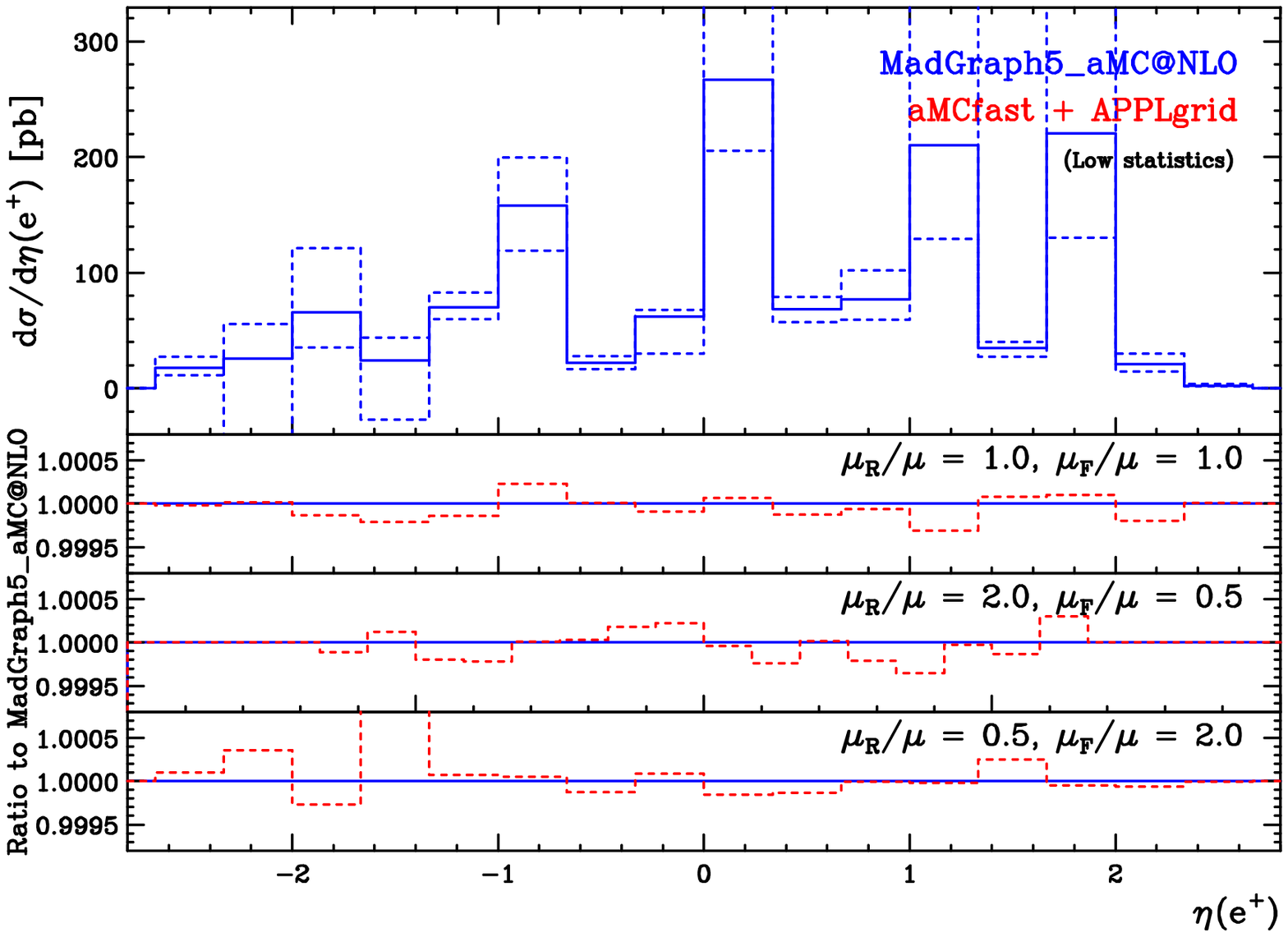, width=0.48\textwidth}
 \epsfig{file=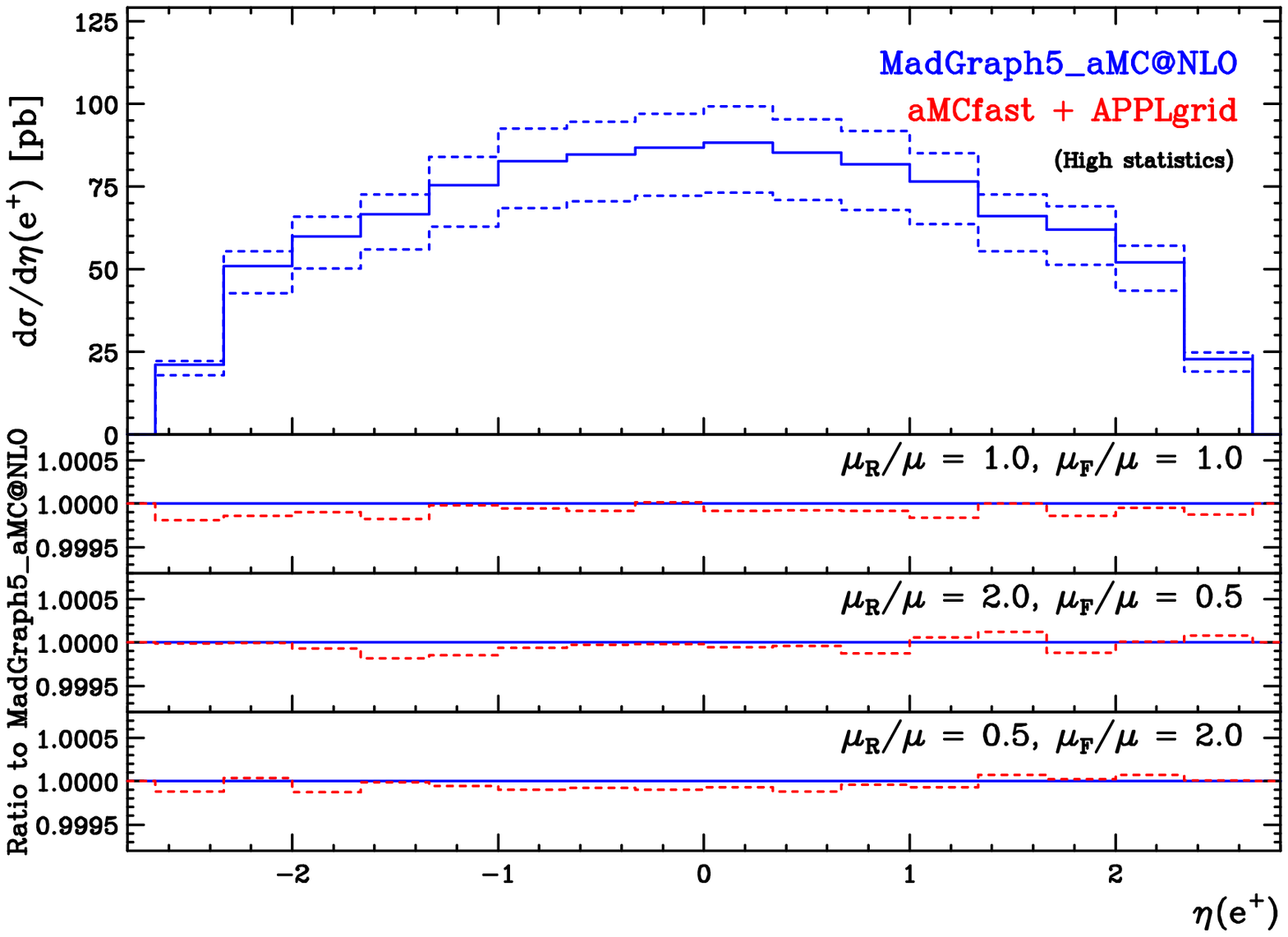, width=0.48\textwidth}
 \end{center}
\caption{As in fig.~\ref{fig:ytt_ttbar}, for the $e^+$ pseudorapidity, in 
$W^+\bar{c}$ hadroproduction.
}
\label{fig:yw_wcharm}
\end{figure}

\begin{figure}
 \begin{center}
 \epsfig{file=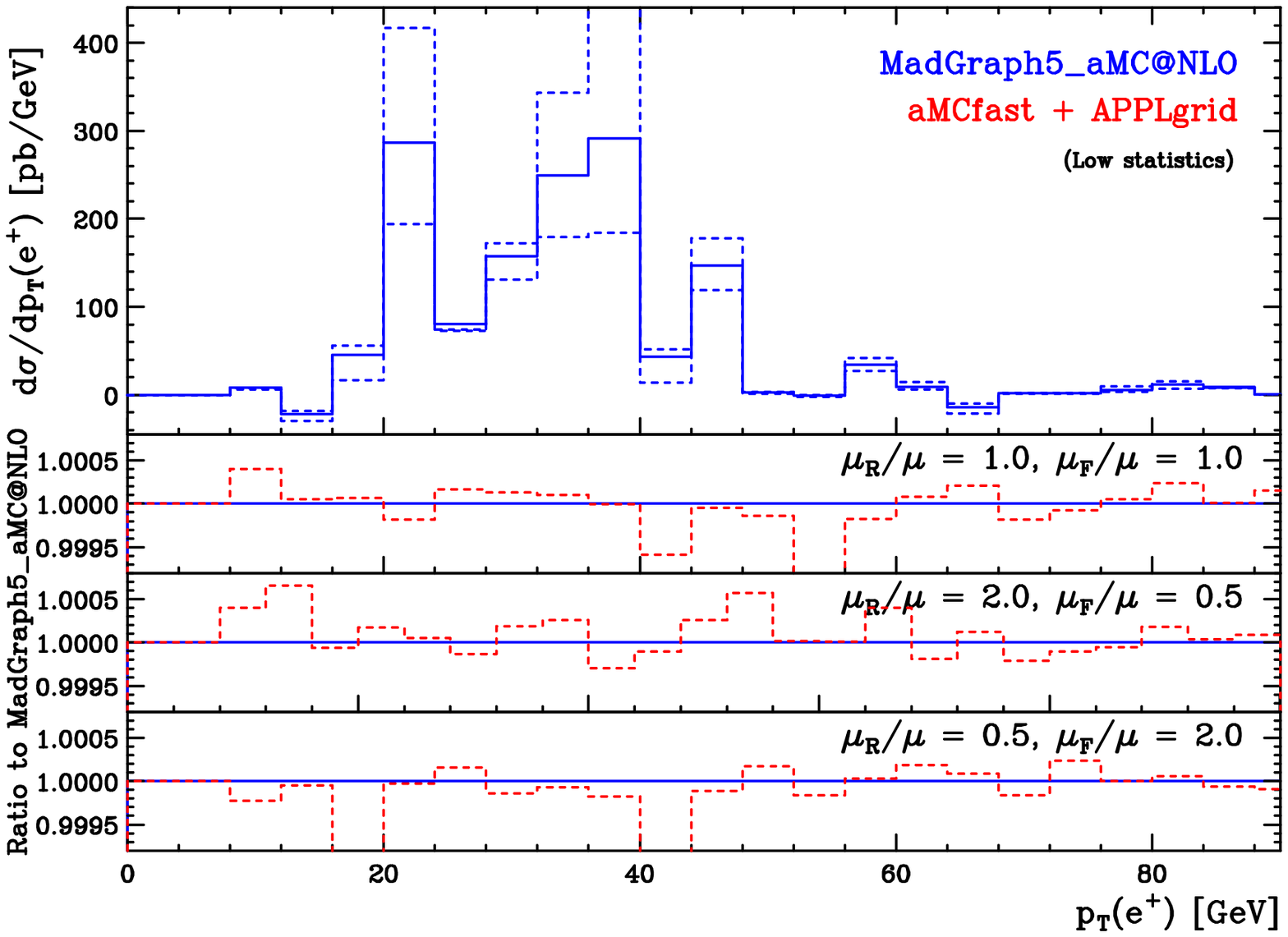, width=0.48\textwidth}
 \epsfig{file=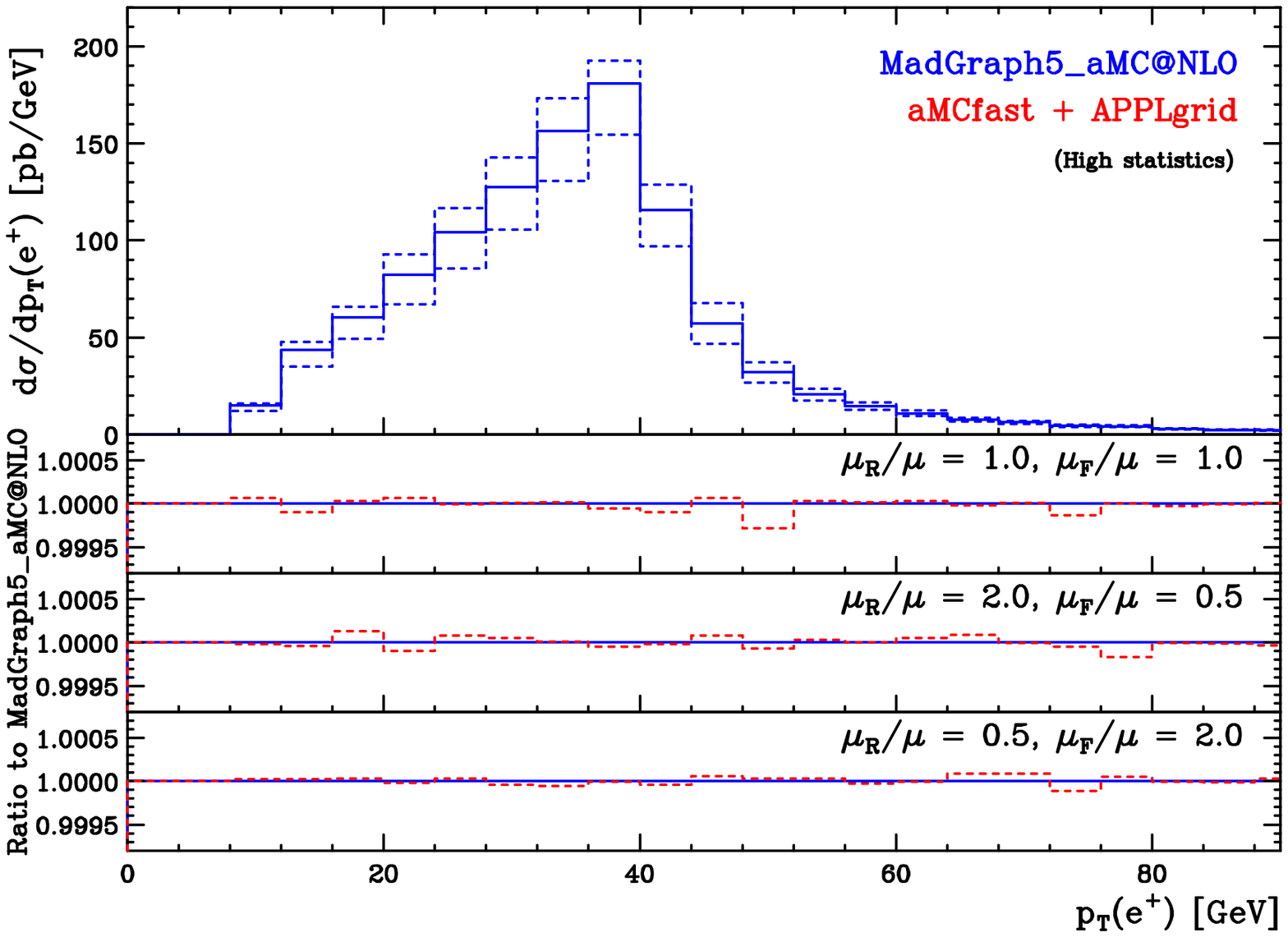, width=0.48\textwidth}
 \end{center}
\caption{As in fig.~\ref{fig:yw_wcharm}, for the $e^+$ transverse momentum.
}
\label{fig:ptw_wcharm}
\end{figure}

The two representative distributions that we use for validation are
the pseudorapidity of the positron, ($\eta(e^+)$, presented 
in fig.~\ref{fig:yw_wcharm}), and its transverse momentum
($\pt(e^+)$ , presented in fig.~\ref{fig:ptw_wcharm}).
The same conclusions as in all previous cases apply here.

\section{Conclusions and outlook\label{sec:outlook}}
In this paper we have presented \amcfast, a tool that serves to achieve
the complete automation of fast NLO QCD computations for arbitrary
processes, by constructing a bridge between \aNLO, which performs
the core automated cross section calculations, and \applgr, which
allows one to represent a given observable in term of interpolating
grids that can be used for a fast and a-posteriori evaluation,
and one in which the possibility is given of changing the scales
and PDFs w.r.t.~those used in the original computation.
Since fast calculations of NLO cross sections are an absolute
necessity in the context of global PDF fits, the combination
of the automated programs discussed in this paper covers all the
present and, more importantly, future needs relevant to
PDF analyses.

We have validated our approach by considering a representative
sample of production processes at the LHC (some of which not
previously available in the form of fast computations), and by comparing
for several observables the bin-by-bin agreement between the 
original results of \aNLO, and those reconstructed by means of
\amcfast\ and \applgr. The typical accuracy that we find
is of the order of some parts in $10^{-4}$, which is in fact
far more than sufficient for any kind of phenomenological application.
Moreover, our approach
allows one to consider scale variations without any loss of
precision, and without the need for interfacing \applgr\
with an external program to perfom PDF evolution.

A natural outcome of this work is the use of the grids produced using
\amcfast\ in actual
global PDF analyses, in view of the possible inclusion into the latter
of LHC data, and especially of those which have not been considered
so far, for lack of either sufficiently precise measurements, or 
of theoretical NLO calculations suited to the task. Given that the format
of the results of the \amcfast\ -- \applgr\ interface is a
trivial extension of that produced by \applgr\ interfaced with
other codes, \amcfast\ should be straighforward to employ by
the global-PDF fitters already experienced with the former code.

The outlook for the near future features two important developments.

Firstly, there is no conceptual difference between what has been
done here, and its analogue in the case of a perturbative expansion
in either the electroweak coupling constant $\aem$, or simultaneously
in $\as$ and $\aem$ (mixed-coupling expansion). From the technical 
viewpoint, the only implication of a mixed-coupling expansion is
the necessity of introducing additional interpolating grids, which
is trivial. This point is important in view of the fact that \aNLO\
is expected to be able to handle soon both kind
of computations; thus, any extension of the \aNLO\ capabilities
will be made immediately available for fast computations, thanks
to straightforward modifications to \amcfast. This will pave the
way to a systematic inclusion of electroweak NLO corrections
into global PDF fits; cases where such an inclusion may prove
particularly relevant include high-$\pt$ jet production, the high-mass
Drell-Yan cross section, as well as the consistent treatment of 
photon-initiated processes that are necessary for the determination
of the photon PDF~\cite{Ball:2013hta}. 
Secondly, the scale- and PDF-independent coefficients 
computed by \aNLO, and used by \amcfast\ to set up the interpolating 
grids at fixed order, are quite similar to those relevant to the
MC@NLO short-distance cross sections. Therefore, the current structure
of \amcfast\ will need only a minor upgrade to be able to deal with
NLO+PS computations. Apart from its obvious phenomenological spinoffs,
and the possibility of obtaining PDF fits that include resummation
effects well tuned to experimental data, such a project is also
interesting from a more theoretical viewpoint, in that it would help
spur a thorough investigation of the effects of the PDFs in initial-state
showers; there is some evidence that these effects are small, but 
more systematic studies should be carried out, in view of the fact
that they cannot be taken into account by any method based solely
on short-distance cross section information.

\noindent
The public version of the \amcfast\ code will soon be available at:\\
$\phantom{aaa}${\tt http://amcfast.hepforge.org/}\\
Meanwhile, potential users are kindly requested to contact the authors.

\section*{Acknowledgments}
We are grateful to Tancredi Carli and
Pavel Starovoitov for discussions on the
\applgr\ framework.
The work of J.~R. has been
partially supported by an STFC Rutherford Fellowship ST/K005227/1.
The work of V.~B. has been supported by the ERC grant 291377, 
``LHCtheory: Theoretical predictions and analyses of LHC physics: 
advancing the precision frontier''.

\appendix
\section{Luminosity factors\label{sec:lum}}
In this appendix, we report the luminosity indices $l$ and parton
identities $(r,s)$ for which the quantity $T_{rs}^{(l)}$ that
enters eq.~(\ref{lumldef}) is equal to one. We do so for all of
the processes studied in sect.~\ref{sec:pheno}, except for $t\bt$ production,
whose analogous information have already been given in sect.~\ref{sec:pheno}.

\begin{table}
{\footnotesize
\begin{center}
\begin{tabular}{cccccccc}
$l$  & $n_{rs}$  & $(r,s)$ & & & & & \\
1 &2 &    $(g \,, u)$&    $(g  \,,c)$ & &  & & \\                                                                                
2 &6 &    $(\bb  \,,u)$  &  $(\bb \,, c)$   & $(\bs  \,,u)$  & $( \bs \,, c)$  &  $(\db \,, u)$ &   $(\db \,, c)$\\              
3 &6 &    $(d  \,,u)$  &  $(d \,, c)$ &   $(s  \,,u)$  & $( s \,, c)$ &  $( b \,, u)$  & $( b \,, c)$\\                          
4 &2 &    $(\cb  \,,c)$   & $(\ub \,, u)$\\                                                                                      
5 &2 &    $(u  \,,u)$   & $(c \,, c)$\\                                                                                          
6 &2 &    $(\cb \,, u)$  &  $(\ub \,, c)$\\                                                                                      
7 &2 &    $(u  \,,c)$   & $(c \,, u)$\\                                                                                          
8 &1 &    $(g  \,,g)$ & \\                                                                                                       
9 &3 &    $(g  \,,d)$  &  $(g \,, s)$ &   $(g \,, b)$\\                                                                          
10 &3 &   $( \bb  \,,b)$  &  $(\bs \,, s)$ &  $( \db \,, d)$\\                                                                   
11 &3 &   $(d  \,,d)$  &  $(s  \,,s)$  &  $(b \,, b)$\\                                                                          
12 &6 &   $(\cb  \,,d)$  &  $(\cb \,, s)$ &   $(\cb  \,,b)$  &  $(\ub \,, d)$ &   $(\ub \,, s)$ &   $(\ub \,, b)$\\              
13 &6 &   $(u  \,,d)$  &  $(u  \,,s)$  &  $(u \,, b)$  &  $(c \,, d)$   & $(c  \,,s)$   & $(c \,, b)$\\                          
14 &6 &   $(\bb \,, d)$   & $(\bb \,, s)$  &  $(\bs \,, d)$   & $(\bs \,, b)$  &  $(\db  \,,s)$   & $(\db \,, b)$\\              
15 &6 &   $(d  \,,s)$  &  $(d \,, b)$  &  $(s  \,,d)$  &  $(s \,, b)$&    $(b \,, d)$  &  $(b  \,,s)$\\                          
16 &2 &   $(g \,, \cb)$  &  $(g  \,,\ub)$\\                                                                                      
17 &6 &   $(\bb \,, \cb)$   & $(\bb  \,,\ub)$   & $(\bs  \,,\cb)$  &  $(\bs  \,,\ub)$  &  $(\db \,, \cb)$  &  $(\db \,, \ub)$\\  
18 &6 &   $(d  \,,\cb)$  &  $(d  \,,\ub)$   & $(s \,, \cb)$  &  $(s  \,,\ub)$ &   $(b  \,,\cb)$ &   $(b  \,,\ub)$\\              
19 &2 &   $(\cb  \,,\cb)$ &   $(\ub \,,\ub)$\\                                                                                   
20 &2 &   $(u  \,,\ub)$  &  $(c  \,,\cb)$\\                                                                                      
21 &2 &   $(\cb  \,,\ub)$  &  $(\ub  \,,\cb)$\\                                                                                  
22 &2 &   $(u \,, \cb)$  &  $(c  \,,\ub)$\\                                                                                      
23 &3 &   $(g  \,,\bb)$  &  $(g \,, \bs)$  &  $(g  \,,\db)$\\                                                                    
24 &3 &   $(\bb  \,,\bb)$  &  $(\bs  \,,\bs)$  &  $(\db  \,,\db)$\\                                                              
25 &3 &   $(d  \,,\db)$  &  $(s  \,,\bs)$  &  $(b  \,,\bb)$\\                                                                    
26 &6 &   $(\cb \,, \bb)$  &  $(\cb  \,,\bs)$  &  $(\cb  \,,\db)$  &  $(\ub \,, \bb)$  &  $(\ub  \,,\bs)$ &   $(\ub \,, \db)$\\  
27 &6 &   $(u  \,,\bb)$   & $(u  \,,\bs)$  &  $(u  \,,\db)$ &   $(c  \,,\bb)$  &  $(c  \,,\bs)$ &   $(c  \,,\db)$\\              
28 &6 &   $(\bb  \,,\bs)$  &  $(\bb  \,,\db)$  &  $(\bs \,, \bb)$ &  $( \bs \,, \db)$  &  $(\db  \,,\bb)$ &   $(\db  \,,\bs)$\\   
29 &6 &   $(d  \,,\bb)$   & $(d  \,,\bs)$ &   $(s  \,,\bb)$  &  $(s  \,,\db)$ &   $(b \,, \bs)$  &  $(b  \,,\db)$\\              
30 &2 &   $(u  \,,g)$   & $(c \,, g)$&\\                                                                                         
31 &3 &   $(d  \,,g)$   & $(s \,, g)$ &   $(b  \,,g)$\\                                                                          
32 &2 &   $(\cb  \,,g)$  &  $(\ub  \,,g)$\\                                                                                      
33 &3 &   $(\bb  \,,g)$  &  $(\bs \,, g)$ &   $(\db \,,g)$\\                                                                     
\end{tabular}
\end{center}
}
\caption{\label{pjlist}
As in table~\ref{ttlist}, for $\gamma+$jet and dilepton$+$jet
production.
}
\end{table}
\begin{table}
{\footnotesize
\begin{center}
\begin{tabular}{cccc}
$l$  & $n_{rs}$  & $(r,s)$ &  \\
1 &1   &   $(g\,, g)$ & \\                 
2 &2   &   $(\bs \,,g)$  &  $(\db \,,g)$ \\
3 &2   &   $(d \,,g )$  & $(s \,,g)$ \\    
4 &2   &   $(\cb \,,g)$  &  $(\ub \,,g)$ \\
5 &2   &   $(u \,,g )$  & $(c \,,g)$ \\    
6 &2   &   $(g \,,\bs)$  &  $(g \,,\db)$ \\
7 &2   &   $(g \,,d )$  & $(g \,,s)$ \\    
8 &2   &   $(g \,,\cb)$  &  $(g \,,\ub)$ \\
9 &2   &   $(g \,,u)$   & $(g \,,c)$ \\    
10& 2  &   $(u \,,\ub)$  &  $(c \,,\cb)$ \\
11 &2  &   $(d \,,\db)$  &  $(s \,,\bs)$ \\
12 &2  &   $(\cb \,,c)$  &  $(\ub \,,u)$ \\
13 &2  &   $(\bs \,,s)$  &  $(\db \,,d)$ \\
\end{tabular}
\end{center}
}
\caption{\label{zbblist}
As in table~\ref{ttlist}, for $Zb\bb$ production.
}
\end{table}
\begin{table}
{\footnotesize
\begin{center}
\begin{tabular}{cccc}
$l$  & $n_{rs}$  & $(r,s)$ &  \\
1 & 1 &   $(g \,, \bs )$& \\                   
2 & 2  &  $(u \,,\bs)$  &  $(d \,, \bs)$ \\    
3 & 2  &  $(\db \,,\bs)$  &  $(\ub \,,\bs)$ \\ 
4 & 1  &  $(s \,,\bs)$ & \\                    
5 & 1  &  $(\bs \,,\bs)$ & \\                  
6 & 1  &  $(g \,,g)$ & \\                      
7 & 1  &  $(\bs \,,g)$ & \\                    
8 & 2  &  $(\bs \,,u)$  &  $(\bs \,,d)$ \\     
9 & 2  &  $(\bs \,,\db)$  &  $(\bs \,,\ub)$ \\ 
10 & 1 &  $(\bs \,,s)$ & \\                    
\end{tabular}
\end{center}
}
\caption{\label{wclist}
As in table~\ref{ttlist}, for $e^+\nu_e \bar{c}$ production.
}
\end{table}

\newpage
\input{aMCfast2.bbl}

\end{document}

%% file: aMCfast2.bbl
\providecommand{\href}[2]{#2}\begingroup\raggedright\endgroup